\pdfoutput=1
\documentclass[twocolumn]{aastex61}

\received{xxx}
\revised{yyy}
\accepted{zzz}

\submitjournal{AAS Journals}

\shorttitle{Tidal Torques in Stellar Binaries}
\shortauthors{Fleming et al.}

\usepackage{hyperref}
\usepackage{xspace}
\usepackage{graphicx}
\usepackage{amsmath}
\usepackage[caption=false]{subfig}

\def\gsim{~\rlap{$>$}{\lower 1.0ex\hbox{$\sim$}}}
\def\lsim{~\rlap{$<$}{\lower 1.0ex\hbox{$\sim$}}}

\newcommand{\vplanet}[0]{\texttt{VPLanet}\xspace}
\newcommand{\eqtide}[0]{\texttt{EQTIDE}\xspace}
\newcommand{\stellar}[0]{\texttt{STELLAR}\xspace}
\newcommand{\kepler}[0]{\textit{Kepler}\xspace}

\begin{document}

\title{Rotation Period Evolution in Low-Mass Binary Stars: The Impact of Tidal Torques and Magnetic Braking}



\author{David P. Fleming}
\affil{Astronomy Department, University of Washington \\
Box 951580, Seattle, WA 98195}
\affil{NASA Astrobiology Institute - Virtual Planetary Laboratory Lead Team, USA}

\author{Rory Barnes}
\affiliation{Astronomy Department, University of Washington \\
Box 951580, Seattle, WA 98195}
\affil{NASA Astrobiology Institute - Virtual Planetary Laboratory Lead Team, USA}

\author{James R. A. Davenport}
\affiliation{Astronomy Department, University of Washington \\
Box 951580, Seattle, WA 98195}

\author{Rodrigo Luger}
\affiliation{Center for Computational Astrophysics, Flatiron Institute \\
New York, NY 10010}
\affil{NASA Astrobiology Institute - Virtual Planetary Laboratory Lead Team, USA}


\begin{abstract}

We examine how tides, stellar evolution, and magnetic braking shape the rotation period (P$_{rot}$) evolution of low-mass stellar binaries up to orbital periods (P$_{orb}$) of 100 d across a wide range tidal dissipation parameters using two common equilibrium tidal models. We find that many binaries with P$_{orb} \lsim 20$ d tidally lock, and most with $P_{orb} \lsim 4$ d tidally lock into synchronous rotation on circularized orbits. At short P$_{orb}$, tidal torques produce a population of fast rotators that single-star only models of magnetic braking fail to produce.  In many cases, we show that the competition between magnetic braking and tides produces a population of subsynchronous rotators that persists for Gyrs, even in short P$_{orb}$ binaries, qualitatively reproducing the subsynchronous eclipsing binaries (EBs) discovered in the \kepler field by \citet{Lurie2017}. Both equilibrium tidal models predict that binaries can tidally-interact out to P$_{orb} \approx 80$ d, while the Constant Phase Lag tidal model predicts that binaries can tidally lock out to P$_{orb} \approx 100$ d. Tidal torques often force the P$_{rot}$ evolution of stellar binaries to depart from the long-term magnetic braking-driven spin down experienced by single stars, revealing that P$_{rot}$ is not be a valid proxy for age in all cases, i.e. gyrochronology can underpredict ages by up to $300\%$ unless one accounts for binarity. We suggest that accurate determinations of orbital eccentricties and P$_{rot}$ can be used to discriminate between which equilibrium tidal models best describes tidal interactions in low-mass binary stars.


\end{abstract}




\section{Introduction} \label{sec:intro}

The long-term angular momentum evolution of low-mass (M$\lsim 1$ M$_{\odot}$) stars is controlled by magnetic braking, the torque exerted on stars due to the coupling of stellar winds to the surface magnetic field \citep{Mestel1968}. Early in stellar lifetimes, stars spin-up as they contract along the pre-main sequence.  Once stars reach the main sequence, stellar radii remain mostly constant while magnetic braking removes angular momentum from the stars, gradually spinning them down over time \citep{Skumanich1972}. Although the precise details of how magnetic braking operates are not fully known, models of magnetic braking have been used to successfully model the bulk trends of P$_{rot}$ distributions in clusters \citep[e.g. Praesepe, ][]{Reiners2012,Matt2015,Douglas2017} and field stars \citep[e.g. the \kepler field, ][]{Matt2015,vanSaders2018}. Furthermore, the magnetic braking-driven long-term spin-down of stars has been used to estimate stellar ages, a method known as gyrochronology \citep{Skumanich1972,Barnes2003,Barnes2007,Mamajek2008,Barnes2010}, with older stars assumed to have lost more angular momentum due to magnetic braking and therefore rotate more slowly.

In contrast, the angular momentum evolution in low-mass short-period (P$_{orb} \lsim 10$ d) stellar binaries is dominated by tides.  Tidal torques drive secular changes in the binary orbit and stellar spins, eventually circularizing the orbit and synchronizing the stellar spins in the long-term \citep{Counselman1973}. Orbital circularization is ubiquitous for short-period binaries, owing to the tidal torque's strong radius and semi-major axis dependence, with both theoretical \citep[e.g.][]{Zahn1989,Claret1995} and observational \citep[e.g.][]{Meibom2005,Mazeh2008,Lurie2017} studies finding that most binaries with P$_{orb} \lsim 10$ d are circularized. For short-period binaries, tidal torques work quickly on ${\sim}100$ Myr timescales, as \citet{Zahn1989} found that the orbit of solar twin binaries circularize during the stellar pre-main sequence.  Observations by \citet{Meibom2005} support this picture as they find short-period binaries in the ${\sim}150$ Myr old cluster M35 tend to have circular orbits.

Tides impart a significant signature in the long-term angular momentum evolution for binary stars, especially for stellar spins. Tidal torques drive binaries towards the tidally locked state in which the stellar P$_{rot}$ is equal to the equilibrium rotation period (P$_{eq}$) predicted by tidal models, with a familiar example of this effect being spin-orbit synchronization where P$_{rot} = $ P$_{eq} = $ P$_{orb}$.  Tidal-locking occurs much earlier than orbital circularization with the tidal-locking timescale estimated to be $2-3$ orders of magnitude less than the circularization timescale \citep{Zahn1989,Witte2002,Mazeh2008} as there is typically much less angular momentum in stellar spins than the binary orbit. As a result, tidal-locking is expected for binaries with P$_{orb} \lsim$ 20 d \citep[e.g.][]{Levato1974,Meibom2006,Mazeh2008,Zahn2008,Meibom2015}.

In low-mass binaries, both magnetic braking and tidal torques compete to shape the stellar P$_{rot}$ evolution. When tides dominate, in particular at close orbital separations, tides can fix P$_{rot} =$ P$_{orb}$, or more generally P$_{rot} =$ P$_{eq}$ for eccentric orbits. In such situations, magnetic braking still operates, removing angular momentum from each star, forcing tides to compensate for each star's loss of angular momentum by spinning up the stars to maintain the tidally locked equilibrium, removing angular momentum from the orbit, hardening the binary \citep[][]{Verbunt1981,Repetto2014,Fleming2018}. Tides do not win out over magnetic braking in general, however, as magnetic braking can spin-down the stars past the tidally locked state into subsynchronous rotation \citep[P$_{rot} >$ P$_{eq}$, ][]{Habets1989,Zahn1994,Keppens1997}. This behavior seems to be bourne out in nature, as \citet{Lurie2017} discovered a substantial population of subsynchronous short-period binaries in the \kepler field, clustered near P$_{orb}/$P$_{rot}{\approx} 0.9$, in defiance of the expectation of tidal locking at such short orbital separations. The competition between magnetic braking and tidal torques can lead to complex angular momentum evolution in low-mass stellar binaries, and no previous work has conducted a systematic study to examine how this evolution proceeds across a wide range of tidal dissipation parameters and P$_{orb}$.

 
Understanding the interaction between tidal torques and magnetic braking is of paramount importance as P$_{rot}$ distributions measured in clusters \citep[e.g. Praesepe, ][]{Agueros2011,Douglas2017} and field stars \citep[e.g. \kepler, ][]{Reinhold2013,McQuillan2014} are likely contaminated by unresolved binaries given that roughly half of Sun-like stars are in stellar binaries \citep{Raghavan2010,Duchene2013}, and that binaries are difficult to resolve in photometric surveys. In the \kepler field, for example, \citet{Simonian2018} recently found that most rapid rotators (P$_{rot} \lsim 7.5$ d) are likely non-eclipsing, tidally-synchronized short-period photometric binaries, indicating that tidal torques in binaries can significantly impact observed P$_{rot}$ distributions.  Tidally-interacting binaries impart a contaminating signal that is not currently accounted for by models. Moreover, any ages inferred from rotation periods of stars in unresolved binaries using gyrochronology could be incorrect owing to the influence of tidal torques. No previous study has quantified this effect. 

There is currently a large number of \kepler binaries with known P$_{rot}$ and P$_{orb}$ \citep[e.g.][]{Lurie2017}. Both the extended \kepler mission \citep[K2,][]{Howell2014} and the Transiting Exoplanet Survey Satellite \citep[TESS, ][]{Ricker2014,Sullivan2015} are expected to detect additional low-mass eclipsing binaries, with Gaia parallaxes \citep{Gaia2016} poised to help refine these stellar parameters, potentially creating a rich dataset of the angular momentum budgets of low-mass binaries. Developing a framework for the angular momentum evolution of low-mass binaries can enable the characterization of the nature of tidal torques in binaries by conditioning on datasets of the spin and orbital states of stellar binaries.

Here, we present a model for the angular momentum evolution of low-mass stellar binaries over their full premain and main sequence lifetimes using a realistic treatment of stellar evolution, magnetic braking, and tidal torques. We investigate under what conditions tidal-locking occurs, and how tidal torques influence rotation in stellar binaries as a function of binary P$_{orb}$ and tidal dissipation parameters for two widely-used equilibrium tidal models.  We show how tidal torques can impact stellar rotation in binaries out to P$_{orb} = 100$ d, causing stellar rotation periods to not strongly correlate with age, making the predictions of gyrochronology models fail in such systems.  We describe our model in $\S$~\ref{sec:methods} and our simulation procedure in $\S$~\ref{sec:simulations}.  We discuss our results in $\S$~\ref{sec:results}, apply our model to the \kepler field in $\S$~\ref{sec:kepler}, and discuss our results' implications in $\S$~\ref{sec:discussion}.


\section{Methods} \label{sec:methods}

We simulate coupled stellar-tidal evolution for low-mass binaries using an improved version of the model presented in \citet{Fleming2018}.  We implement our model in the open-source code VPLanet\footnote{VPLanet is publicly available
at \href{https://github.com/VirtualPlanetaryLaboratory/vplanet}{{https://github.com/VirtualPlanetaryLaboratory/vplanet}}.} \citep{Barnes2019}.  We integrate all model equations (see $\S$~\ref{sec:methods:stellar} and $\S$~\ref{sec:methods:eqtide}) using the $4^{th}$ order Runge-Kutta scheme with adaptive timestepping described in \citet{Fleming2018}.  

\subsection{Stellar Evolution} \label{sec:methods:stellar}

We improve upon the interpolation of the \citet{Baraffe2015} stellar evolution models employed by \citet{Fleming2018}, \stellar, by additionally performing a bicubic interpolation of the stellar radius of gyration, $r_g$, over mass and time of the \citet{Baraffe2015} models. This updated version of \stellar now tracks the full moment of inertia evolution of low-mass stars according to the \citet{Baraffe2015} stellar evolution models, a critical requirement for modeling the angular momentum evolution of low-mass stars. 

We simulate magnetic braking using the model derived by \citet{Matt2015} as this formalism has been shown to successfully model the spin-down of low-mass stars across many ages in both the Praesepe cluster and in the \kepler field. This model depends on the stellar Rossby number, $Ro = $ P$_{rot}/\tau_{cz}$, the ratio of the stellar P$_{rot}$ to the stellar convective turnover timescale, $\tau_{cz}$. The \citet{Matt2015} model predicts that below a certain $Ro$ for rapidly-rotating stars, stellar magnetic activity saturates at a constant value, producing a magnetic braking torque that is directly proportional to the stellar rotation rate.  The angular momentum loss for rapidly-rotating saturated stars is given by
\begin{equation} \label{eqn:mattSat}
\frac{dJ}{dt} = -\frac{dJ}{dt}\Bigg|_0 \chi^2 \left( \frac{\omega}{\omega_{\odot}} \right) 
\end{equation}
while for more slowly-rotating unsaturated stars,
\begin{equation} \label{eqn:mattUnSat}
\frac{dJ}{dt} = -\frac{dJ}{dt}\Bigg|_0 \left( \frac{\tau_{cz}}{\tau_{cz \odot}} \right)^2 \left( \frac{\omega}{\omega_{\odot}} \right)^3
\end{equation}
where
\begin{equation} \label{eqn:matt0}
\frac{dJ}{dt}\Bigg|_0 = 6.3 \times 10^{30} \ \mathrm{erg} \ \left( \frac{R}{R_{\odot}} \right)^{3.1} \left( \frac{M}{M_{\odot}} \right)^{0.5}.
\end{equation}
Saturated magnetic braking occurs for $Ro \leq Ro_{\odot}/\chi$ for $\chi = 10$ where \citet{Matt2015} defines $\chi = Ro_{\odot}/Ro_{sat}$.  We adopt all model parameters given in Table 1 from \citet{Matt2015}, with the correction from \citet{Matt2019}, and compute $\tau_{cz}$ using Eqn. (36) from \citet{Cranmer2011}.


We model the net change in the stellar rotation rate due to stellar evolution and magnetic braking via the following equation 
\begin{equation} \label{eqn:stellar_rot_rate_dt}
\dot{\omega} = \frac{\dot{J}_{mb}}{I} - \frac{2 \dot{R} \omega}{R} - \frac{2 \dot{r_g} \omega}{r_g}
\end{equation}
where the moment of inertia $I = M r_g^2 R^2$, $\dot{J}_{mb}$ is the angular momentum loss due to magnetic braking, and the time derivatives of the stellar $R$ and $r_g$ are computed numerically using our interpolation of the \citet{Baraffe2015} stellar evolution grids.  

\subsubsection{Core-Envelope Coupling} \label{sec:methods:twoLayer}

Our simplified model assumes that stars follow solid body rotation, whereas in real low-mass stars, coupling between the radiative core and convective envelope can impact the surface rotation period evolution \citep{MacGregor1991,Allain1998,Bouvier2008,Irwin2009}. Recent work by \citet{Gallet2013} and \citet{Gallet2015} find that the rapidly-rotating stellar core acts as an angular momentum reservoir for the convective envelope, potentially transferring angular momentum within the stellar interior and into the envelope for up to 1 Gyr, depending on the adopted magnetic braking model, the initial rotation rate, and the stellar mass. We anticipate that internal angular momentum transport would work against the spin-down caused by tidal torques, increasing tidal locking timescales. Internal angular momentum transport torques could potentially balance both tidal and magnetic braking torques near the tidally locked state, producing slight supersynchronous rotation, analogous to the subsynchronous case examined in $\S$~\ref{sec:eq}. Modeling core-envelope coupling is beyond the scope of this work, however.

\subsubsection{Example Stellar Evolution} \label{sec:methods:exampleStellar}

\begin{figure*}[ht]
	\includegraphics[width=\textwidth]{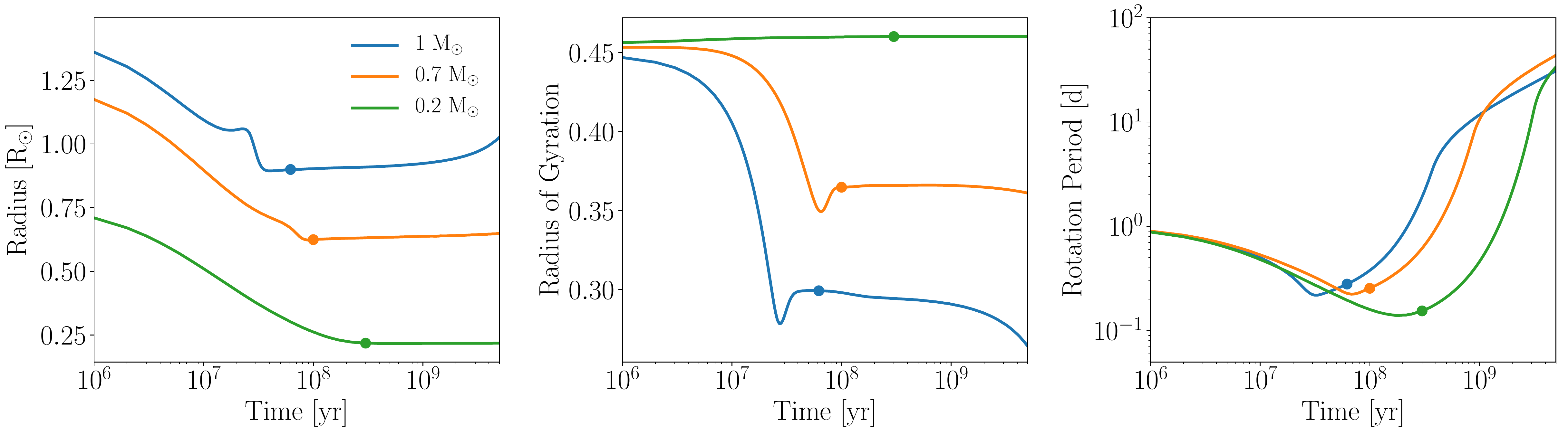}
   \caption{Stellar $R$ (left), $r_g$ (middle), and P$_{rot}$ (right) evolution for 0.2 M$_{\odot}$ (M, green), 0.7 M$_{\odot}$ (K, orange), and 1 M$_{\odot}$ (G, blue) mass stars computed according to \stellar, our interpolation of the \citet{Baraffe2015} stellar evolution models ($\S$~\ref{sec:methods:stellar}) combined with the \citet{Matt2015} magnetic braking model. Each dot denotes the approximate time when each star reaches the main sequence.}%
    \label{fig:stellarExample}%
\end{figure*}

In Fig.~\ref{fig:stellarExample}, we plot the evolution of $R$, $r_g$, and P$_{rot}$ for 0.2 M$_{\odot}$, 0.7 M$_{\odot}$, and 1 M$_{\odot}$ mass stars, representing an M, K , and G dwarf, respectively, computed according to \stellar and the \citet{Matt2015} magnetic braking model. We assume all stars have an initial P$_{rot} = 1$ d and have an initial age of 5 Myr. All stars' radii contract along the pre-main sequence, spinning the stars up (right panel). Once the stars reach the main sequence, their structure changes slowly, allowing magnetic braking to dominate the stellar angular momentum evolution, significantly spinning-down the stars over long timescales. The $r_g$ evolution noticeably differs between the stars as the late M dwarf's (green) $r_g$ varies little as it remains fully convective, while the K and G dwarf grow a radiative core while on the pre-main sequence, decreasing $r_g$ until both reach the main sequence.


\subsection{Tidal Evolution} \label{sec:methods:eqtide}

 Equilibrium tidal models, first introduced by \citet{Darwin1880}, track the secular evolution of an orbiter's semi-major axis, $a$, eccentricity, $e$, and the rotation rates, $\omega_i$, and obliquities $\psi_i$, of both gravitating bodies due to tidal torques. Equilibrium tidal models assume that tidally interacting bodies raise tidal bulges on their companions that remain offset from the line connecting the bodies' centers of mass due to friction within each body. This assumption is typically referred to as the ``weak friction approximation" \citep{Zahn2008}.  The tidal bulges cause torques that permit the exchange of angular momentum between the orbit and both bodies' spins. Equilibrium tidal models are linear since they assume that the tidal waves that comprise the tidal bulge raised on a body are uncoupled. Under these assumptions, the tidal evolution is analogous to a driven, damped harmonic oscillator \citep{Greenberg2009}. For low-mass stars, equilibrium tidal models assume that tidal forces primarily dissipate energy in the outer-convective regions via viscous turbulence \citep[see][]{Zahn2008}. Although simple, equilibrium tidal models have been used to model the secular orbital and rotation evolution of both Solar System bodies and exoplanets \citep[e.g.][]{Goldreich1966,Jackson2009,Leconte2010,Heller2011,Barnes2013,Barnes2017} and stellar binaries \citep[e.g.][]{Zahn1989,Zahn2008,Khaliullin2011,Repetto2014,Fleming2018}. We refer the reader to \citet{Barnes2017} for an in-depth discussion of the assumptions and limitations of equilibrium tidal models. Here, we consider two common equilibrium tidal models to study the secular spin-orbital evolution of low-mass stellar binaries.  

\subsubsection{Constant Phase Lag Model}

The ``Constant Phase Lag" (CPL) \citep[][]{FerrazMello2008,Heller2011} equilibrium tidal model assumes that the tidal torque on one body due to its companion arises from a linear combination of several discrete, uncoupled tidal bulges, each with its own associated frequency, that maintain a fixed phase offset with respect to the line connecting the two stars' centers of mass. We use the \eqtide implementation of the CPL model in \vplanet following the derivation of \citet{FerrazMello2008}.  The equations that govern the secular change in $e$ and $a$ are as follows:

\begin{equation} \label{eqn:cpl:e}
\frac{de}{dt} = -\frac{ae}{8 G m_1 m_2} \sum_{i=1}^2 Z_{i,\mathrm{CPL}} \left( 2 \varepsilon_{0,i} - \frac{49}{2} \varepsilon_{1,i} + \frac{1}{2} \varepsilon_{2,i} + 3 \varepsilon_{5,i} \right)
\end{equation}
\begin{equation} \label{eqn:cpl:a}
\frac{da}{dt} = \sum_{i=1}^2 \frac{da_i}{dt}
\end{equation}
where if the $i^{th}$ body is tidally locked in a synchronous orbit,
\begin{equation} \label{eqn:cpl:dadt_locked}
\frac{da_{i,sync}}{dt} = -\frac{a^2}{G m_1 m_2} Z_{i,\mathrm{CPL}} \left( 7 e^2 + \sin^2 (\psi_i) \right) \varepsilon_{2,i},
\end{equation}
otherwise
\begin{equation}
\begin{split}
\frac{da_i}{dt} & = \frac{a^2}{4 G m_1 m_2} Z_{i,\mathrm{CPL}} \left( 4 \varepsilon_{0,i} + e^2 \left[ -20 \varepsilon_{0,i} + \frac{147}{2} \varepsilon_{1,i} \right. \right. \\
&  + \left. \left. \frac{1}{2} \varepsilon_{2,i} - 3 \varepsilon_{5,i} \right] - 4 \sin^2 (\psi_i) \left[ \varepsilon_{0,i} - \varepsilon_{8,i} \right] \right).
\end{split}
\end{equation}
The CPL equations for $\psi$ and $\omega$ evolution are
\begin{equation} \label{eqn:cpl:psi}
\frac{d\psi_i}{dt} = \frac{Z_{i,\mathrm{CPL}} \sin(\psi_i)}{4 m_i r_{g,i}^2 R_i^2 n \omega_i} \left( [1-\xi_i] \varepsilon_{0,i} + [1+\xi_i](\varepsilon_{8,i} - \varepsilon_{9,i}) \right)
\end{equation}
\begin{equation} \label{eqn:cpl:omega}
\begin{split}
\frac{d\omega_i}{dt}& = -\frac{Z_{i,\mathrm{CPL}}}{8m_i r_{g,i}^2 R_i^2 n} \left(4 \varepsilon_{0,i} + e^2\left[-20\varepsilon_{0,i} + 49\varepsilon_{1,i} + \varepsilon_{2,i} \right] \right. \\
& \left. + 2 \sin^2(\psi_i) \left[ -2 \varepsilon_{0,i} + \varepsilon_{8,i} + \varepsilon_{9,i} \right] \right)
\end{split}
\end{equation}
where $G$ is Newton's gravitational constant, $n$ is the binary's mean motion, and the index $i$ denotes that $i^{th}$ body. The tidal phase lags signs, $\varepsilon$, for the $i^{th}$ body are given by
\begin{equation} \label{eqn:cpl:eps}
\begin{split}
\varepsilon_{0,i} & = \Sigma(2 \omega_i - 2n) \\
\varepsilon_{1,i} & = \Sigma(2 \omega_i - 3n) \\
\varepsilon_{2,i} & = \Sigma(2 \omega_i - n) \\
\varepsilon_{5,i} & = \Sigma(n) \\
\varepsilon_{8,i} & = \Sigma(\omega_i - 2n) \\
\varepsilon_{9,i} & = \Sigma(\omega_i)
\end{split}
\end{equation}
where the function $\Sigma(x)$ returns $1$ for positive $x$, $-1$ for negative $x$, and $0$ otherwise.

The intermediate variable $Z_{\mathrm{CPL},i}$ is given by
\begin{equation} \label{eqn:cpl:z}
Z_{i,\mathrm{CPL}} = 3 G^2 k_{2,i} M_j^2 (M_i + M_j) \frac{R_i^5}{a^9} \frac{1}{n Q_i}
\end{equation}
where the $j^{th}$ body is the $i^{th}$ body's companion, $k_{2}$ is the body's Love number of degree 2, and $Q$ is the tidal quality factor (``tidal Q"). The tidal Q parameterizes the energy dissipation due to tidal evolution, with lower tidal Qs, i.e. larger phase differences between the tidal bulges, driving more rapid tidal evolution.

The other intermediate variable, $\xi_i$, is defined as
\begin{equation}\label{eqn:cpl:chi}
\xi_i = \frac{r_{\mathrm{g},i}^2 R_i^2 \omega_i a n }{ G M_j}.
\end{equation}

\subsubsection{Constant Time Lag Model}

The ``Constant Time Lag" (CTL) \citep[][]{Hut1981,Leconte2010} equilibrium tidal model assumes a constant time interval between the body's tidal bulge and the passage of the tidally interacting companion. In this formalism, unlike the CPL model, the CTL model is continuous over a range of tidal wave frequencies and applicable for large $e$.  However, if the assumption of linearity is relaxed, i.e. frequencies associated with tidal bulges are allowed to depend on a spin or orbital forcing frequency, then this model is only valid over a small range of frequencies \citep{Greenberg2009}. We use the \eqtide implementation of the CTL model in \vplanet following the derivation of \citet{Leconte2010}.  The equations that govern the secular changes in $e$, $a$, $\omega$, and $\psi$ are as follows:

\begin{equation} \label{eqn:ctl:e}
  \frac{de}{dt} = \frac{11 ae}{2 G M_1 M_2}
  \sum_{i = 1}^2 Z_{\mathrm{CTL},i} \left( \cos(\psi_i) \frac{f_4(e)}{\beta^{10}(e)}  \frac{\omega_i}{n} -\frac{18}{11} \frac{f_3(e)}{\beta^{13}(e)}\right),
\end{equation}

\small
\begin{equation}\label{eqn:ctl:a}
  \frac{da}{dt} \ = \  \frac{2 a^2}{G M_1 M_2}
  \sum\limits_{i = 1}^2 Z_{\mathrm{CTL},i} \left( \cos(\psi_i) \frac{f_2(e)}{\beta^{12}(e)} \frac{\omega_i}{n} - \frac{f_1(e)}{\beta^{15}(e)}\right),
\end{equation}

\begin{equation}\label{eqn:ctl:omega}
  \frac{d\omega_i}{dt} \ = \ \frac{Z_{\mathrm{CTL},i}}{2 M_i r_{g,i}^2 
R_i^2 n} \left( 2 \cos(\psi_i) \frac{f_2(e)}{\beta^{12}(e)} - \left[ 1+\cos^2(\psi)
 \right] \frac{f_5(e)}{\beta^9(e)} 
\frac{\omega_i}{n} \right),  
\end{equation}
and
\begin{equation}\label{eqn:ctl:psi}
  \frac{d\psi_i}{dt} = \frac{Z_{\mathrm{CTL},i} \sin(\psi_i)}{2 M_i r_{g,i}^2 R_i^2 n \omega_i}\left( \left[ \cos(\psi_i) - \frac{\xi_i}{ \beta} \right] \frac{f_5(e)}{\beta^9(e)} \frac{\omega_i}{n} - 2 \frac{f_2(e)}{\beta^{12}(e)} \right).
\end{equation}
\normalsize
where the intermediate variables are given by 
\begin{equation}\label{eqn:ctl:z}
 Z_{i,\mathrm{CTL}} = 3 G^2 k_{2,i} M_j^2 (M_i+M_j) \frac{R_i^5}{a^9} \tau_i ,
\end{equation}
and 
\begin{equation}\label{eqn:ctl:f_e}
\begin{array}{l}
\beta(e) = \sqrt{1-e^2},\\
f_1(e) = 1 + \frac{31}{2} e^2 + \frac{255}{8} e^4 + \frac{185}{16} e^6 + \frac{25}{
64} e^8,\\
f_2(e) = 1 + \frac{15}{2} e^2 + \frac{45}{8} e^4 + \frac{5}{16} e^6,\\
f_3(e) = 1 + \frac{15}{4} e^2 + \frac{15}{8} e^4 + \frac{5}{64} e^6,\\
f_4(e) = 1 + \frac{3}{2} e^2 + \frac{1}{8} e^4,\\
f_5(e) = 1 + 3 e^2 + \frac{3}{8} e^4.
\end{array}
\end{equation}

In both the CPL and CTL model, We assume $k_2 = 0.5$. This choice of $k_2$ does not impact our results as $k_2$ is degenerate with Q in the CPL model, e.g. the $k_2/Q$ scaling in Eq.~(\ref{eqn:cpl:z}), and with $\tau$ in the CTL model, e.g. $k_2 \tau$ scaling in Eq.~(\ref{eqn:ctl:z}), so we instead examine how our results scale with $Q$ and $\tau$.  Any constraints we derive as a function $Q$ or $\tau$ can trivially be scaled to other values of $k_2$. For example, a common re-parameterization of Q is the reduced tidal quality factor, $Q' = 3Q/2k_2$ \citep[e.g.][]{Leconte2010}. Given our choice of $k_2 = 0.5$, this reduces to $Q' = 3 Q$.

\subsubsection{Tidal Locking}

Tidal torques drive a body's rotation rate towards the tidally locked state. When a body tidally locks, tidal torques fix P$_{rot}$ to the equilibrium P$_{rot}$, P$_{eq}$.  Typically, tidal locking is understood in the context of a synchronized rotator, e.g. when P$_{rot} = $ P$_{eq} = $ P$_{orb}$. Although spin-orbit synchronization is an expected outcome of tidal evolution \citep{Counselman1973}, in general for tidally locked bodies on non-circular orbits, both the CPL and CTL model predict pseudosynchronous, or supersynchronous rotation, e.g. Mercury's 3:2 spin-orbit resonance \citep[P$_{rot} = 2/3$ P$_{orb}$,][]{GoldreichPeale1966}.   

The CPL model, owing to its assumption of a finite number of discrete tidal lags, only permits a 1:1 and 3:2 spin-orbit state where, following \citet{Barnes2017}, the CPL P$_{eq}$ is given by
\begin{equation} \label{eqn:cpl:eqPer}
P^{\mathrm{CPL}}_{eq} = 
\begin{cases}
P_{orb} & \text{if } e < \sqrt{1/19}\\
\frac{2}{3}P_{orb} & \text{if } e \geq \sqrt{1/19}.
\end{cases}
\end{equation}
Therefore, the CPL model predicts synchronous rotation for $e \lsim 0.23$, and a supersychronous 3:2 spin-orbit state otherwise for tidally locked rotators.

We note that two discrete rotation states are not the only permitted ones for tidally locked systems under the CPL formalism. For example, an alternate derivation of P$_{eq}$ for orbiters with rotation axes perpendicular to the orbital plane under the CPL model predicts
\begin{equation} \label{eqn:cpl:eqPerCont}
P_{eq} = \frac{P_{orb}}{1 + 9.5e^2},
\end{equation}
a continous function of $e$ \citep{Goldreich1966b,Murray1999}. Here, we follow the suggestions of both \citet{Barnes2013} and \citet{Barnes2017} and use the discrete P$_{eq}$ version of the CPL model for self-consistency.

The CTL model is continuous over a range of tidal frequencies and therefore predicts a P$_{eq}$ that is a continuous function of both $e$ and $\psi$.  Following \citet{Barnes2017}, we define the CTL P$_{eq}$ by
\begin{equation} \label{eqn:ctl:eqPer}
P^{\mathrm{CTL}}_{eq} = P_{orb} \frac{\beta^3 f_5(e) (1 + \cos^2(\psi))}{2f_2(e) \cos(\psi)}.
\end{equation}
The CTL model predicts that bodies on eccentric orbits tidally lock into supersyncronous rotation, and only bodies with aligned spins on circular orbits are synchronous rotators. 

In general, a continuous P$_{eq}$ and the discrete 1:1 and 3:2 spin-orbit commensurabilities are not the only equilibrium rotation states for tidally locked rotators predicted by equilibrium tidal models. For example, \citet{Rodriguez2012} show that tidally interacting bodies can get captured into many spin-orbit resonances states, e.g. 2:1, 5:2, 4:3, etc, and below, we search for evidence of them in data of the spin-orbital states of \kepler EBs. Note that our model does not resolve capture into such states as the CPL model, owing to its inclusion of only 4 discrete tidal lags, only allows a body to enter into 3:2 and 1:1 spin - orbit commensurabilities. The CTL model predicts a continuous equilibrium period as a function of the P$_{orb}$, $e$, and obliquity, for tidally locked bodies, only resolving capture into 1:1 synchronous rotation.

\subsubsection{Numerical Details of Tidal Locking}

Due to the discontinuities in the equilibrium tidal model equations, for example in Eq.~(\ref{eqn:cpl:eps}) when $\omega \approx n$, and due to the inherant discreteness of numerical integrations, numerical solutions for the CPL and CTL models can produce unphysical evolution. We follow \citet{Barnes2013} and \citet{Fleming2018} and fix P$_{rot} = $ P$_{eq}$ according to Eq.~(\ref{eqn:cpl:eqPer}) or Eq.~(\ref{eqn:ctl:eqPer}) for the CPL and CTL models, respectively, when P$_{rot}$ is within $1\%$ of P$_{eq}$.  To ensure that tidal torques dominate over torques due to magnetic braking and stellar evolution when forcing tidal-locking, we additionally require that the P$_{rot}$ derivative points towards $P_{eq}$ on both sides of P$_{eq}$, i.e. when the gradient of P$_{rot}$ points towards the tidally locked state, before fixing P$_{rot} = $ P$_{eq}$. We find that this scheme produces physically and numerically accurate results. 

\subsubsection{The Dynamical Tide} \label{sec:methods:dynamicalTide}

An additional mechanism for tidal dissipation in low-mass stellar binaries is the dynamical tide. This effect arises from the turbulent viscous damping of inertial waves that are excited in the stellar convective envelope by a tidal perturber, with Coriolis acceleration serving as the restoring force \citep{Zahn1975,Ogilvie2007}. Under the dynamical tide formalism, the stellar mass, evolving stellar structure, rotation rate, and tidal forcing frequency can all strongly impact the strength of tidal dissipation, which can span many orders of magnitude \citep{Ogilvie2007,Ogilvie2013,Mathis2015,Gallet2017}. For example, adopting the tidal frequency-averaged model for tidal dissipation of \citet{Ogilvie2013}, both \citet{Mathis2015} and \citet{Gallet2017} show that dynamical tidal dissipation is enhanced during the pre-main sequence due to the expansion of the stellar radiative core and rapid rotation, whereas the magnetic braking-driven spin-down on the main sequence decreases the tidal dissipation. Dissipation due to the dynamical tide could be important for some of the systems considered in this work since, for binary stars on circular orbits, inertial waves are excited in the stellar convective envelopes for P$_{orb} > $P$_{rot}/2$ and can drive significant spin and orbital evolution \citep[e.g.][]{Witte2002,Ogilvie2007,Bolmont2016}. Although semi-analytic models for dynamical tidal dissipation that account for the evolving stellar structure and rotation exist \citep[e.g.][]{Mathis2015,Bolmont2016,Gallet2017}, we do not consider them here as they are currently limited to circular orbits. We instead focus on exploring the impact of the equilibrium tide across a wide range of parameter space and leave an examination of how the combination of the dynamical and equilibrium tide impacts the rotation period evolution of low-mass binary stars for future work.

\subsubsection{Example Tidal Evolution} \label{sec:methods:eqtideExample}

\begin{figure}
	\includegraphics[width=0.4\textwidth]{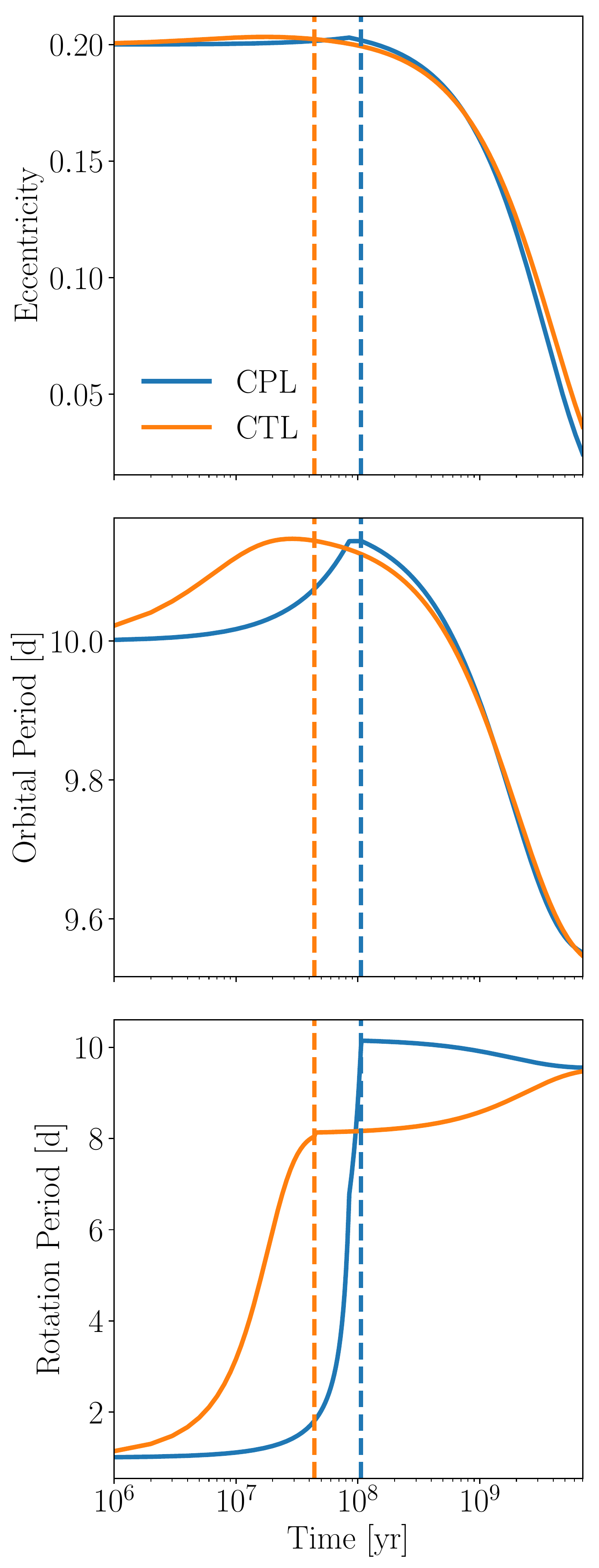}
   \caption{Tidal evolution of a 1 M$_{\odot} -$ 1 M$_{\odot}$ stellar binary's $e$ (top), P$_{orb}$ (middle), and P$_{rot}$ (bottom) for the CPL (blue) and CTL (orange) model. The blue (CPL) and orange (CTL) vertical dashed lines denote when the stellar binary tidally locks. Both the CPL and CTL model predict the same qualitative evolution. The rotational evolution differs, however, as under the CPL model, the binary tidally locks into a synchronous orbit as $e < \sqrt{1/19}$, e.g. Eq.~(\ref{eqn:cpl:eqPer}), while the CTL model predicts supersyncronous rotation due to the CTL model's equilibrium period eccentricity dependence (see Eq.~(\ref{eqn:ctl:eqPer})).}%
    \label{fig:tidalExample}%
\end{figure}

We plot the tidal evolution for $a$, $e$, and $P_{rot}$, ignoring stellar evolution, for a solar-twin binary with an initial P$_{orb} = 10$ d, P$_{rot} = 1$ d, $e = 0.2$ for the CPL model and CTL model, assuming $Q=10^6$ and $\tau = 0.1$ seconds, respectively, in Fig.~\ref{fig:tidalExample}. The CPL and CTL model predict the same qualitative evolution: both the binary's $e$ and P$_{orb}$ slightly increase as tides force the spins toward the tidally locked state, transferring rotational angular momentum into the orbit, increasing the orbital angular momentum by ${\sim}1\%$ in the process.  At late times, both the CPL and CTL drive the binaries towards orbital circularization, with tidal dissipation decreasing P$_{orb}$. The predictions of the CPL and CTL model, differ, however, when the binaries tidally lock.  Under the CPL model, the binary tidally locks into a synchronous orbit when $e < \sqrt{1/19}$, e.g. Eq.~(\ref{eqn:cpl:eqPer}), while the CTL model predicts supersyncronous rotation due to the CTL model's equilibrium period eccentricity dependence, e.g. Eq.~(\ref{eqn:ctl:eqPer}).

\subsection{Coupled Stellar-Tidal Evolution For tidally locked Systems} \label{sec:coupled}

Following \citet{Fleming2018}, when one or both binary stars are tidally locked, tidal forces prevent magnetic braking from spinning down the tidally locked star(s), and any angular momentum lost comes at the expense of the binary orbit, decreasing $a$ as a result \citep{Verbunt1981}.  Below in Eq.~(\ref{eqn:tidal_locked_one}) and Eq.~(\ref{eqn:tidal_locked_two}), we modify the $a$ decay equations due to stellar evolution and magnetic braking in tidally locked binaries from \citet{Fleming2018}, their Eqs. (18) and (20), to additionally account for $r_g$ evolution when one or both stars tidally lock, respectively, assuming conservation of angular momentum:
\small
\begin{equation} \label{eqn:tidal_locked_one}
\begin{split}
\dot{a}_{coupled}^{(1)} = \frac{-\dot{J}_{mb} - 2 \omega \left( m_1 r_{g,1}^2 R_1 \dot{R_1} - m_1 r_{g,1} \dot{r}_{g,1} R_1^2 \right)}
{\frac{\mu^2 G M (1-e^2)}{2J_{orb}} - \frac{3 \omega}{2a} m_1 r_{g,1}^2 R_1^2}
\end{split}
\end{equation}
\normalsize
and
\small
\begin{equation} \label{eqn:tidal_locked_two}
\begin{split}
\dot{a}_{coupled}^{(2)} = \frac{-\dot{J}_{mb} - 2 \omega \left( \sum_{i=1}^{2} m_i r_{g,i}^2 R_i \dot{R_i} + m_i r_{g,i} \dot{r}_{g,i} R_i^2 \right)}
{\frac{\mu^2 G M (1-e^2)}{2J_{orb}} - \frac{3 \omega}{2a} \left( m_1 r_{g,1}^2 R_1^2 + m_2 r_{g,2}^2 R_2^2 \right)},
\end{split}
\end{equation}
\normalsize
where $J_{orb}$ is the orbital angular momentum.

\section{Simulations} \label{sec:simulations}

We examine stellar angular momentum evolution in low-mass binaries by simulating two sets of 10,000 stellar binaries, one modeled using the CPL model and the other using the CTL formalism.  We simulate both stars' spin evolution but mainly consider the P$_{rot}$ evolution for the primary, i.e. more massive star in binaries, as it is observationally easier to measure a P$_{rot}$ on the more massive, and hence brighter, star \citep[e.g.][]{Meibom2006,Lurie2017}. For each simulation, we sample the primary's mass uniformly over $[0.1, 1]$ M$_{\odot}$. Following \citet{Matt2015}, we uniformly sample the log$_{10}$ of P$_{rot}$ over [$0.8,15$] days, a distribution that approximates the P$_{rot}$ distribution of young stars in the ${\sim}2$ Myr old Orion Nebula Cluster \citep{Stassun1999,Herbst2001,Herbst2002,Rodriguez-Ledesma2009}.  We compute the secondary star's mass by uniformly sampling the mass ratio over [$0.1, 1$] following observations of mass ratios in low-mass binaries \citep{Raghavan2010,Moe2018}. Given the inherent uncertainty in and complexity of the formation of short-period binaries \citep[e.g.][]{Bonnell1994,Bate2000,Bate2002,Moe2018} and the potential for dynamical processing via tides or stellar close encounters \citep[e.g.][]{Mardling2001,Hurley2002,Ivanova2005,Meibom2005}, we take an agnostic approach to the initial orbital configuration by uniformly randomly sampling the initial eccentricity ($e$) over [$0.0,0.3$], consistent with eccentricities of field binaries that likely have not been tidally-processed \citep{Raghavan2010}. Although the CTL model is applicable for $e \gsim 0.3$, the CPL model is not and can predict qualitatively incorrect evolution in that regime \citep[see Section 4.1 in][]{Leconte2010}, so we restrict $e \leq 0.3$ to allow us to compare both models. We uniformly sample the initial P$_{orb}$ over [$3,100$] d and do not consider P$_{orb} < 3$ d as these binaries are likely to have a tertiary companion \citep{Tokovinin2006} which can significantly impact the inner binary's dynamical evolution \citep[e.g.][]{Fabrycky2007,Munoz2015,Martin2015b,Hamers2016,Moe2018}. 

Values for stellar tidal $Q$s and $\tau$s for low-mass stars are highly uncertain due to complex viscous evolution within the stars \citep{Ogilvie2007}, and can differ for stars of the same spectral class \citep{Barker2009}. These parameters can also vary as a function of stellar mass or age \citep{Bolmont2016,vanEylen2016}, likely due to low-mass stars' evolving convective regions where the tidal dissipation predominantly occurs \citep{Zahn2008}. Typical values of $Q$ and $\tau$ for Sun-like stars are estimated to be of order $Q \approx 10^6$ and $\tau \approx 0.1$ s, respectively \citep[e.g.][]{Meibom2005,Ogilvie2007,Jackson2008}, however a range of values exist in the literature.  Therefore, we consider a wide range of tidal parameters by sampling stellar tidal $Q$s log-uniformly over $[10^4,10^8]$ and $\tau$ log-uniformly over $[10^{-2},10]$ s.  There is no general expression to compute $Q$ as a function of $\tau$, or vice versa, except in some special cases where approximations exist, e.g. Eqn. (2) from \citet{Heller2011}. All stars have an initial age of 5 Myr unless stated otherwise as by this time, the gaseous protoplanterary circumbinary disk that can drive significant dynamical evolution in the binary \citep[e.g.][]{Fleming2017} would likely have dissipated \citep{Haisch2001}. We also perform a smaller subset of simulations to illustrate the behaviour of our coupled model and describe their initial conditions as we introduce them. All code used to run simulations and generate figures is available online.\footnote{\href{https://github.com/dflemin3/sync}{https://github.com/dflemin3/sync}.}



\section{Results} \label{sec:results}

\subsection{Interaction Between Magnetic and Tidal Braking: Subsynchronous Rotation} \label{sec:eq}

Here we focus on binaries in the ``weak tides" regime, i.e. long P$_{orb}$ and large $Q$ or small $\tau$, to identify the boundary between evolution dominated by tides or magnetic braking via analytic calculations and simulations.  

\subsubsection{Analytic Torque Balance} \label{sec:analytic}

\begin{figure}[h]
	\includegraphics[width=0.5\textwidth]{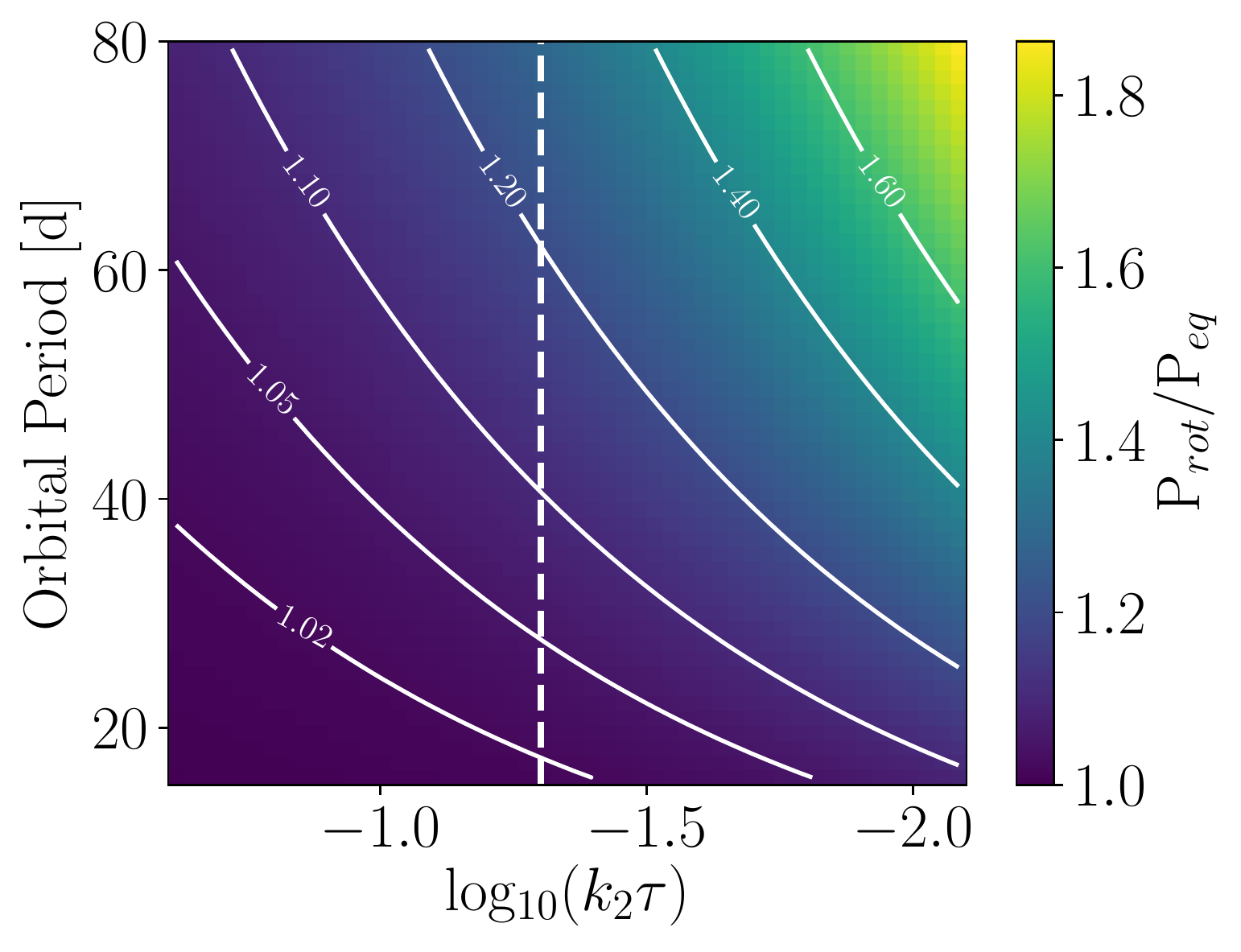}
   \caption{The stellar P$_{rot}$, normalized by P$_{eq}$, at which the torques due to magnetic braking and tides balance for a 1 M$_{\odot} - 1$ M$_{\odot}$ binary on a circular orbit according to Eqn.~\ref{eqn:appendixBalance}. The white dashed line indicates our fiducial values for $k_2$ and $\tau$, 0.5 and 0.1 s, respectively, that we adopt in the simulations in $\S$~\ref{sec:mattBalance}.}%
    \label{fig:analyticBalance}%
\end{figure}

In the weak tides regime, spin-down due to magnetic braking will drive the stellar P$_{rot}$ past P$_{eq}$, resulting in subsynchronous rotation, P$_{rot} >$ P$_{eq}$. For long P$_{orb}$, the stars will be slowly-rotating and in the unsaturated regime \citep{Matt2015}. Since magnetic braking scales as P$_{rot}^{-3}$ for unsaturated rotators, e.g. Eqn.~(\ref{eqn:mattUnSat}), magnetic braking torques weaken as the stellar rotation slows down, so at some P$_{rot}$, tidal torques will balance magnetic braking, producing a long-lasting state of subsynchronous rotation. We compute the P$_{rot}$ at which this balance occurs as a function of P$_{orb}$, $k_2$, and $\tau$ in $\S$~\ref{sec:appendix:balance} by setting the sum of Eqn.~(\ref{eqn:ctl:omega}) and Eqn.~(\ref{eqn:mattUnSat}) equal to 0, considering tidal torques under the CTL formalism. For simplicity, we assume both stars are solar-mass with 0 obliquity, a circular binary orbit, and that the torque balance occurs while the stars are on the main sequence where stellar properties change slowly. Although solar mass stars are the most massive stars we consider in this work, and hence will have the strongest tidal torque for a given tidal dissipation parameter and P$_{orb}$, they can still exhibit subsynchronous rotation and serve as a useful end member case to examine here and in simulations below.  We display the results of this calculation in Fig.~\ref{fig:analyticBalance}, normalizing P$_{rot}$ by P$_{eq}$, which for binary stars with 0 obliquity on circular orbits is simply P$_{orb}$.

Our calculations show that subsynchronous rotation occurs across a wide range of tidal parameters and P$_{orb}$. In general as tides weaken, i.e. increasing P$_{orb}$ and/or decreasing $\log_{10}(k_2 \tau)$, tidal and magnetic braking torques balance at longer P$_{rot}$. For strong tides, $\log_{10}(k_2 \tau) \gsim -1$, tidal torques overpower magnetic braking for P$_{orb} \lsim 40$ d, tidally locking binaries into synchronous rotation. For our fiducial values of $k_2 = 0.5$ and $\tau = 0.1$ s (white dashed line in Fig.~\ref{fig:analyticBalance}), solar-twin binaries will rotate subsynchronously for P$_{orb} \gsim 20$ d, with more severe subsynchronism at longer P$_{orb}$. This simple calculation, however, does not account for stellar evolution or secular tidal orbital evolution, e.g. tidal friction that will shrink the orbit, gradually strengthening tidal torques, so we turn to simulations to characterize this evolution. 


\subsubsection{Torque Balance} \label{sec:mattBalance}

\begin{figure*}
	\includegraphics[width=\textwidth]{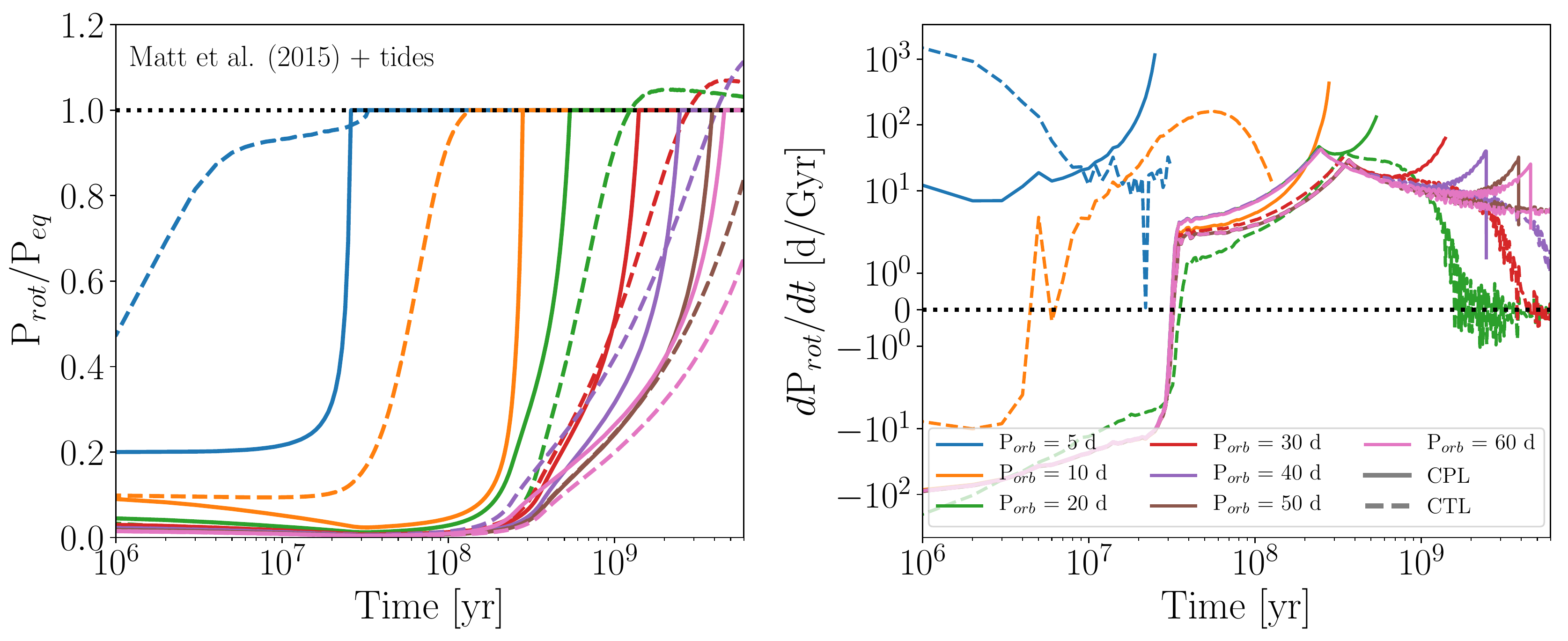}
   \caption{Evolution of stellar P$_{rot}$, normalized by P$_{eq}$ (see Eqn.~(\ref{eqn:cpl:eqPer}) and Eqn.~(\ref{eqn:ctl:eqPer}), for initial circular binary orbits according to the CPL (solid) and CTL (dashed) models with $Q = 10^6$ and $\tau = 0.1$ s, respectively, using the \citet{Matt2015} magnetic braking model. Left: P$_{rot}/$P$_{eq}$ for stars with P$_{orb}$ ranging from 5 d to 60 d. The black dotted line indicates the tidally locked state.  Right: Net P$_{rot}$ derivative due to stellar evolution, tidal torques, and magnetic braking.  We truncate each curve when the binary tidally locks. The legend denotes the initial binary orbital period and we note that the orbital periods do not vary by more than a few percent over the course of the simulations.} For P$_{orb} \leq 10$ d, both the CPL and CTL models predict that the binaries lock into synchronous rotation.  For all P$_{orb}$, the CPL models tidally lock whereas the CTL model predicts subsynchronous rotation that persists for Gyrs.%
    \label{fig:eqPer}%
\end{figure*}

We simulate the full coupled stellar-tidal evolution of 1 M$_{\odot} - 1$ M$_{\odot}$ binaries on initially circular orbits to examine how stellar binaries evolve towards subsynchronous rotation. In Fig.~\ref{fig:eqPer}, we plot P$_{rot}$, normalized by P$_{eq}$, and its time derivative for P$_{orb} \in [5,60]$ d modeled using both the CPL (solid line, $Q=10^6$) and CTL (dashed line, $\tau = 0.1$ s) models. Both tidal models predict that binaries with P$_{orb} < 10$ d will tidally lock within 100 Myr, in agreement with observations \citep{Meibom2005} and previous theoretical work \citep{Zahn1989}. The CPL model predicts that all binaries tidally lock, even out to P$_{orb} = 60$ d, indicating that tidal locking is not necessarily restricted to short P$_{orb}$ systems. As anticipated by our analytic calculations, the CTL model predicts subsynchronous rotation for P$_{orb} \geq 20$ d as magnetic braking overpowers tidal torques. For P$_{orb} = 20$ d, magnetic braking pushes P$_{rot}/$P$_{eq} \approx 1.05$, with the maximum value set by the torque balance. As shown in $\S$~\ref{sec:analytic}, the peak P$_{rot}/$P$_{eq}$ grows for longer P$_{orb}$ since tides weaken with increasing binary separation, e.g. Eqn.~(\ref{eqn:ctl:z}), allowing magnetic braking to dominate the spin evolution. 

Subsynchronous rotation does not persist indefinitely, however, as P$_{rot}$ eventually decreases back towards the tidally locked state in the long-term due to a combination of three simultaneous physical effects.  First, magnetic braking weakens at long P$_{rot}$ as its torque scales as P$_{rot}^{-3}$ for unsaturated rotators \citep{Matt2015}. Second, as P$_{rot}$ increases further from the tidally locked state, tidal torques strengthen as they try to force P$_{rot}$ back towards P$_{eq}$ (see Eqn.~(\ref{eqn:ctl:omega})). Third, when P$_{rot} > $ P$_{eq}$, tides transfer angular momentum from the orbit into stellar rotations, decreasing P$_{orb}$, gradually strengthening tidal torques that strongly depend on the binary separation as $a^{-6.5}$.  These effects combine to shift the balance of power from magnetic braking-controlled stellar spin down to tidal torques spinning-up stars, shepherding them towards P$_{eq}$ in the long-term. 

We can see this process unfold in the right panel of Fig.~\ref{fig:eqPer} where we plot the total P$_{rot}$ time derivative due to tidal torques, stellar evolution, and magnetic braking. Early on, $\dot{\mathrm{P}}_{rot} < 0$ as stars contract along the pre-main sequence until about 60 Myr when the stars reach the zero age main sequence. Tides and magnetic braking then combine to spin down stars towards the tidally locked state. For the CTL models with P$_{orb} > 10$ d, $\dot{\mathrm{P}}_{rot} > 0$ as magnetic braking dominates, driving the stars into subsynchronous rotation. In the long-term, $\ddot{\mathrm{P}}_{rot} < 0$, however, as the three processes described above gradually strengthen tidal torques relative to magnetic braking.  Tidal torques eventually overpower magnetic braking, seen as a slight negative P$_{rot}$ derivative, slowly driving P$_{rot}$ back towards P$_{eq}$, producing a population of subsynchronous rotators that can persist for Gyrs. We explore this point further in $\S$~\ref{sec:shortPorbSubSync}.

\subsection{Influence of P$_{orb}$, $Q$ and $\tau$} \label{sec:qTauMaps}

\begin{figure*}[ht]
	\includegraphics[width=\textwidth]{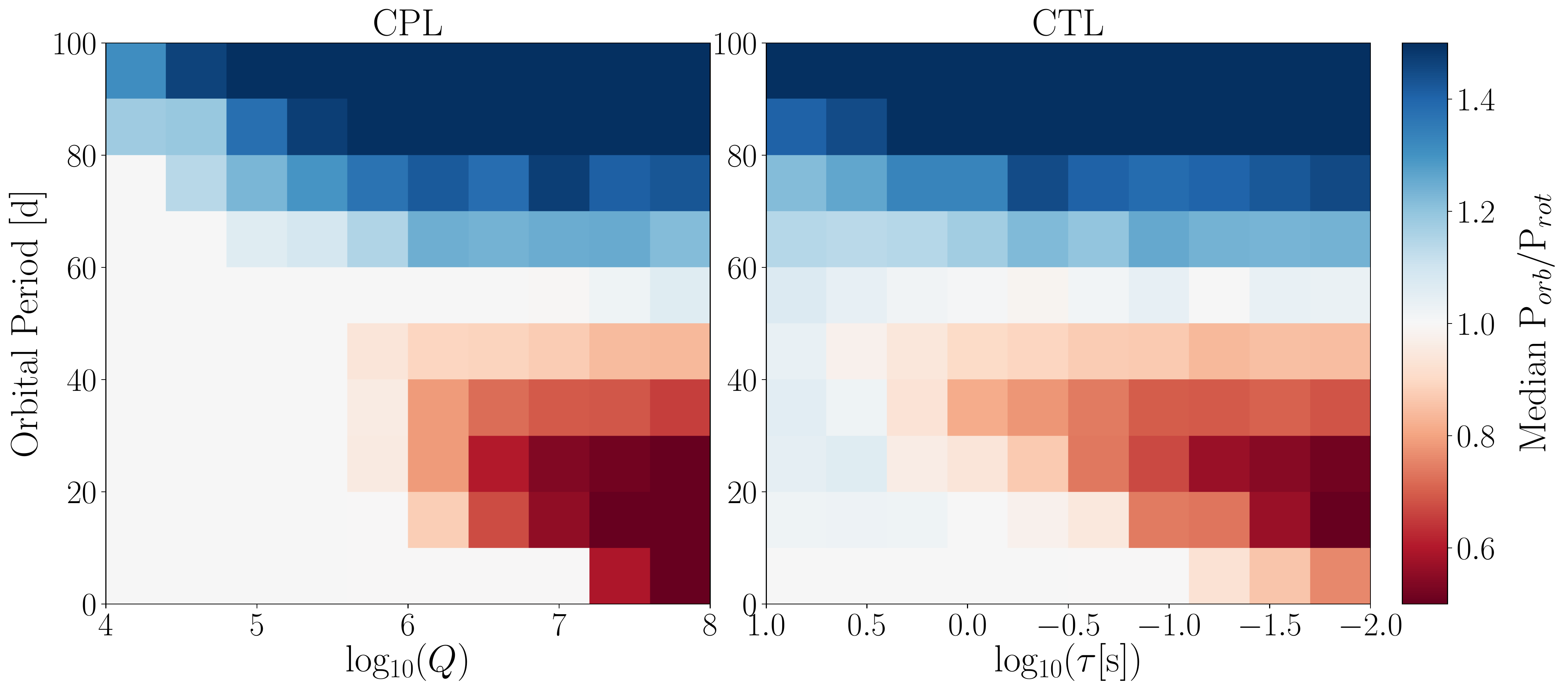}
   \caption{Median P$_{orb}/$P$_{rot}$ at the end of the simulation according to the CPL (left) and CTL (right) models binned by log$_{10}(Q)$ and log$_{10}(\tau)$, respectively, and P$_{orb}$.  For P$_{orb} < 10$ d, the CPL model predicts that most systems will tidally lock into synchronous rotation, whereas the CTL model requires $\tau \geq 0.1$ s to tidally lock.  For large $Q$ ($> 10^7$) and small $\tau$ ($\tau < 0.1$ s), weak tidal torques cannot prevent magnetic braking from spinning down stars past the tidally locked state, producing a population of subsynchronous rotators (red regions, P$_{rot} >$ P$_{orb}$).}%
    \label{fig:qTauLock}%
\end{figure*}

We next examine how P$_{rot}$ evolution in stellar binaries depends on P$_{orb}$ and the strength of tidal dissipation, parameterized by $Q$ and $\tau$ for the CPL and CTL models, respectively. In Fig.~\ref{fig:qTauLock}, we bin our simulation results after the full 7 Gyr evolution by P$_{orb}$ and $Q$ or $\tau$ and compute the median P$_{orb}/$P$_{rot}$ in each bin, marginalizing over all other parameters.

Spin-orbit synchronization is the typical outcome for binaries with P$_{orb} < 10$ d according to the CPL model for most values of $Q$. The strong tidal torques predicted by the CPL model can even tidally lock binaries out to P$_{orb} \gsim 80 $ d for $Q < 10^5$, well beyond the expected limit of 20 d \citep{Meibom2006}.  According to the CTL model, binaries with P$_{orb} < 10$ d typically tidally lock for $\tau \gsim 0.1$ s, and seldomly tidally lock for P$_{orb} > 20$ d, except for systems with strong tides, $\tau \gsim 3$ s.  Both models predict a substantial population of subsynchronous rotators (red regions in Fig.~\ref{fig:qTauLock}, P$_{rot} >$ P$_{orb}$), consistent with magnetic braking dominating weak tidal torques.  The population of supersynchronous rotators (blue regions in Fig.~\ref{fig:qTauLock}, P$_{rot} <$ P$_{orb}$) with P$_{orb} > 60$ d does not in general correspond to binaries tidally locking into supersynchronous rotation, but rather, typically arises from the combination of weak tidal torques and magnetic braking not spinning down stars enough for P$_{rot}$ to be close to the tidally locked state. At a given age, longer P$_{orb}$ binaries will tend to rotate faster as they experience weaker tidal torques, and hence require longer to spin down towards the tidally locked state.

Both tidal models predict a population of nearly synchronous rotators near P$_{orb} \approx 60$ d.  This population corresponds to the evolution described in $\S$~\ref{sec:eq} in which magnetic braking initially spins down stars past the tidally locked state, but in the long-term, tidal torques spin up the stars, shepherding them towards the tidally locked state.  This process can keep stellar P$_{rot} \gsim$ P$_{eq}$ for several Gyrs or longer, depending on the P$_{orb}$ and $Q$ or $\tau$ (see Fig.~\ref{fig:eqPer}, $\S$~\ref{sec:shortPorbSubSync}). 

\begin{figure*}
	\includegraphics[width=\textwidth]{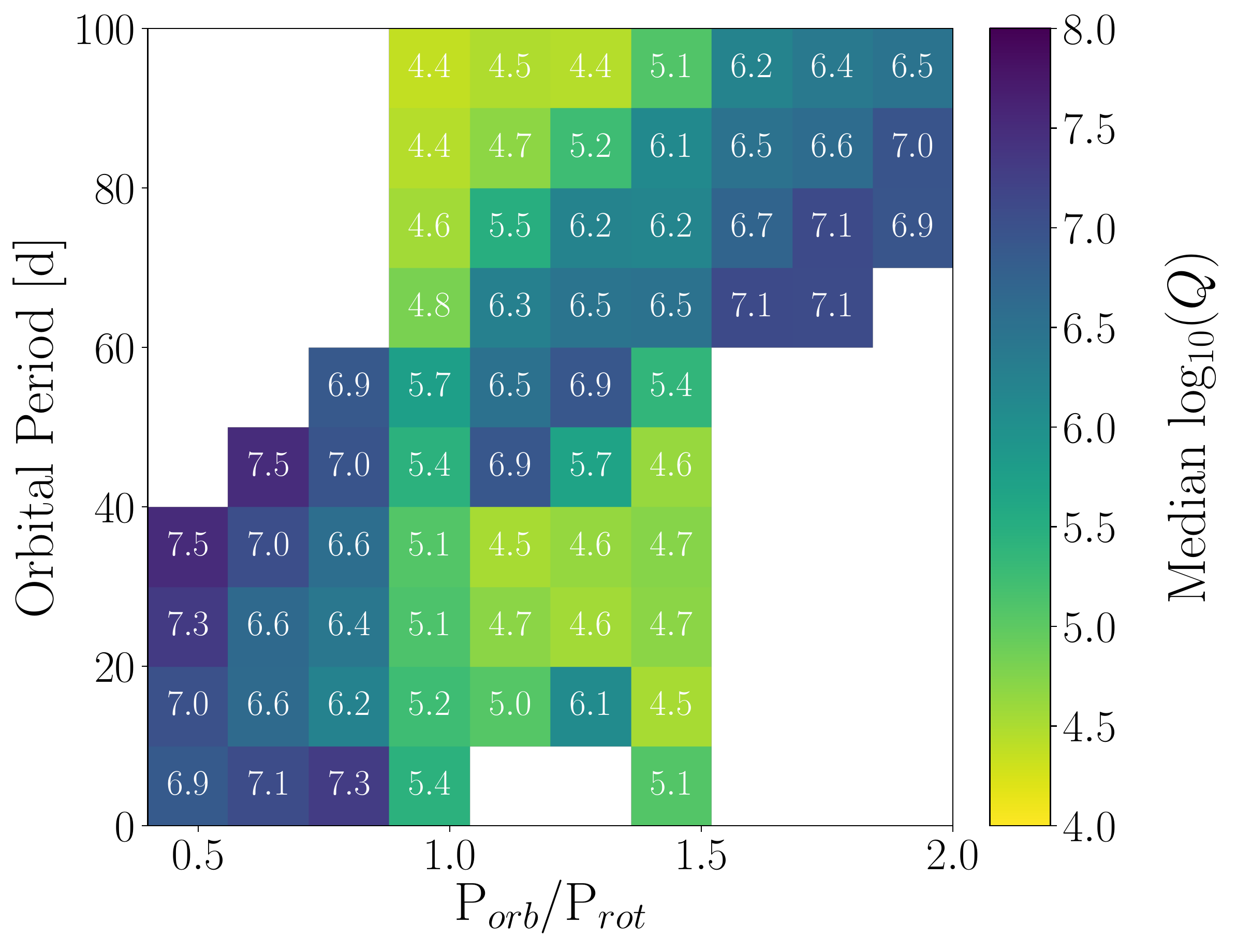}
   \caption{Median log$_{10}(Q)$ of primary stars binned by P$_{orb}$ and P$_{orb}/$P$_{rot}$ evolved using the CPL model. }%
    \label{fig:qmap}%
\end{figure*}

\begin{figure*}
	\includegraphics[width=\textwidth]{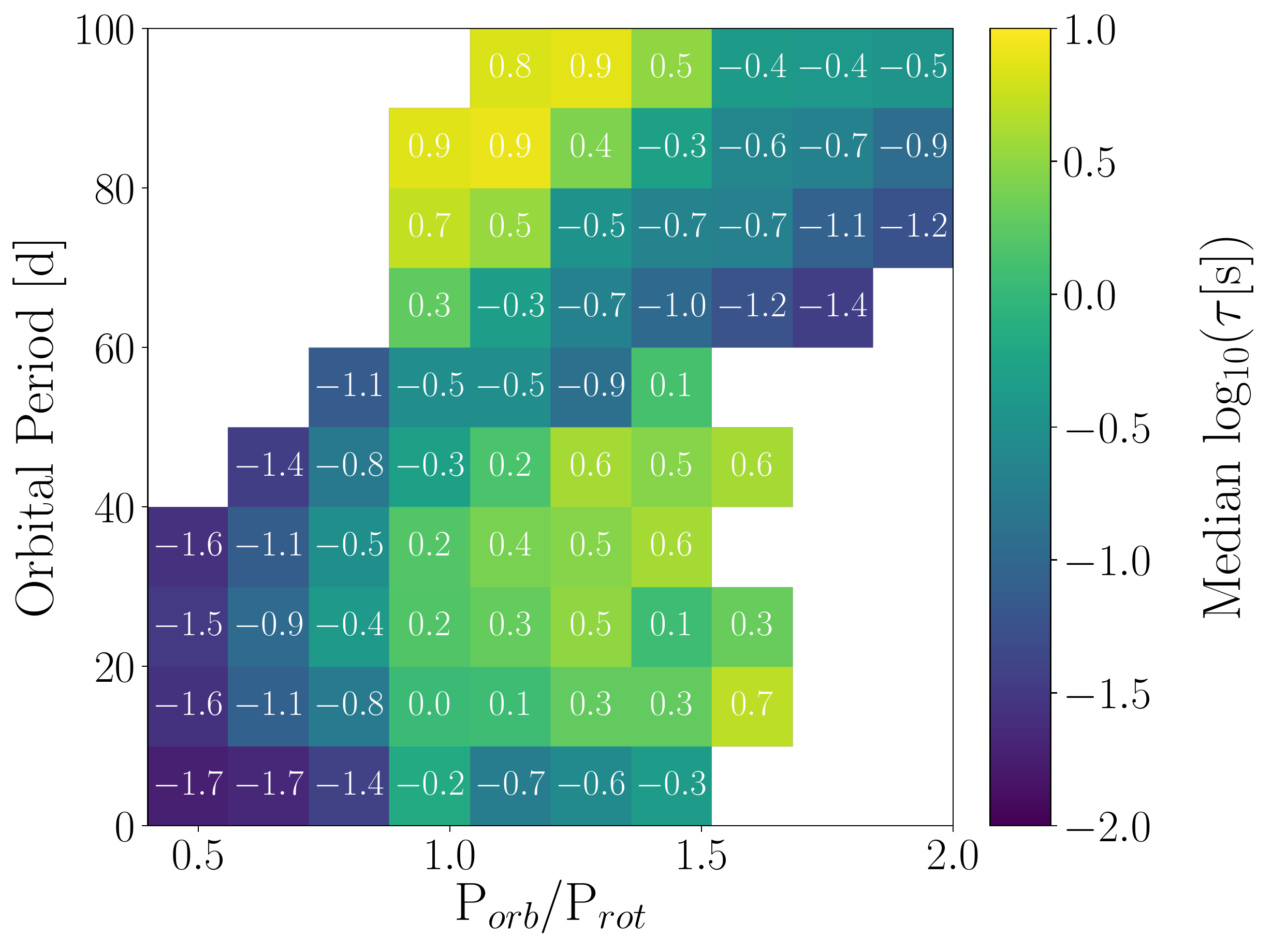}
   \caption{Same format as Fig.~\ref{fig:qmap}, but for log$_{10}(\tau [s])$ under the CTL model. }%
    \label{fig:taumap}%
\end{figure*}

We isolate the impact of $Q$ and $\tau$ on the spin-orbital state of tidally interacting stellar binaries by binning our CPL and CTL simulation results after 7 Gyr of evolution by P$_{orb}$ and P$_{orb}/$P$_{rot}$ in Figures \ref{fig:qmap} and \ref{fig:taumap}, respectively. In these figures, we estimate the typical strength of tidal torques, using $Q$ and $\tau$ as a proxy, that can produce various spin-orbital states. For the CPL simulations depicted in Fig.~\ref{fig:qmap}, synchronous and supersynchronous rotators have systematically low $Q$s, typically $Q < 10^6$, as strong tidal torques are required to tidally lock these binaries. For P$_{orb} < 10$ d, there are no rotators with $1.0 <$ P$_{orb}/$P$_{rot} < 1.5$, nor do any stars have P$_{orb}/$P$_{rot} > 1.5$ for P$_{orb} < 60$ d, as in the CPL model, binaries with eccentric orbits can only tidally lock into a 1:1 or 3:2 spin-orbit commensurability, see Eqn.~(\ref{eqn:cpl:eqPer}).   

Subsynchronous rotators have systematically larger $Q$s, typically $Q > 10^6$, and hence experience weak tidal torques that are dominated by magnetic braking.  Subsynchronous rotation can occur under the CPL model for binaries with P$_{orb} < 50$ d. In this regime, the median $Q$ tends to increase with decreasing P$_{orb}/$P$_{rot}$, except near the tidally locked state, as magnetic braking dominates weaker tidal torques, yielding longer P$_{rot}$.  This trend reverses at longer P$_{orb} > 60$ d where supersynchronous rotation arises from the inability of tidal torques and magnetic braking to spin-down stars enough to approach the tidally locked state by the end of the simulation.  In this case, the more supersynchronous the rotation, the weaker the tidal torques must be, and hence the larger the $Q$ must be. 

According to the CTL model simulations, depicted in Fig.~\ref{fig:taumap}, many binaries tidally lock for P$_{orb} \lsim 20$ d when $\tau \gsim 0.1$ s, with some tidally locking up to P$_{orb} \approx 50$ d when $\tau \gsim 1$ s. Subsynchronous rotation typically occurs for stars with $\tau < 0.1$ s. Similar to the behavior depicted in Fig.~\ref{fig:qmap}, longer P$_{rot}$ are produced by binaries with weaker tidal interactions since P$_{orb}/$P$_{rot}$ decreases monotonically with $\tau$ for P$_{orb} < 40$ d.  For P$_{orb} > 50$ d, magnetic braking dominates the evolution seen in the diagonal sequence with a median $\tau \approx 0.1$ s, a value that is typically insufficient for tides to strongly influence the evolution given the wide orbital separations. The shape of this diagonal region arises from the combination of magnetic braking and our flat initial P$_{orb}$ distribution. In this P$_{orb}$ regime, most binaries rotate supersynchronously as tides and magnetic braking fail to sufficiently spin down the stars by the age of the system. At longer P$_{orb}$, some binaries can strongly tidally-interact, but these systems require extreme tidal $\tau \gsim 10$ s. Unlike the CPL simulations depicted in Fig.~\ref{fig:qmap}, these binaries with P$_{orb} \approx 90$ d are not tidally locked as the tides are still not strong enough to lock the system. We explore this point further in $\S$~\ref{sec:protDist}. We do not often observe P$_{orb}/$P$_{rot} \gsim 1.5$ as we only consider eccentricities up to $e = 0.3$, limiting how rapid supersynchronous systems can rotate according to Eqn.~(\ref{eqn:ctl:eqPer}).


\subsubsection{Subsynchronous Rotation at Short P$_{orb}$} \label{sec:shortPorbSubSync}

\begin{figure}
	\includegraphics[width=0.45\textwidth]{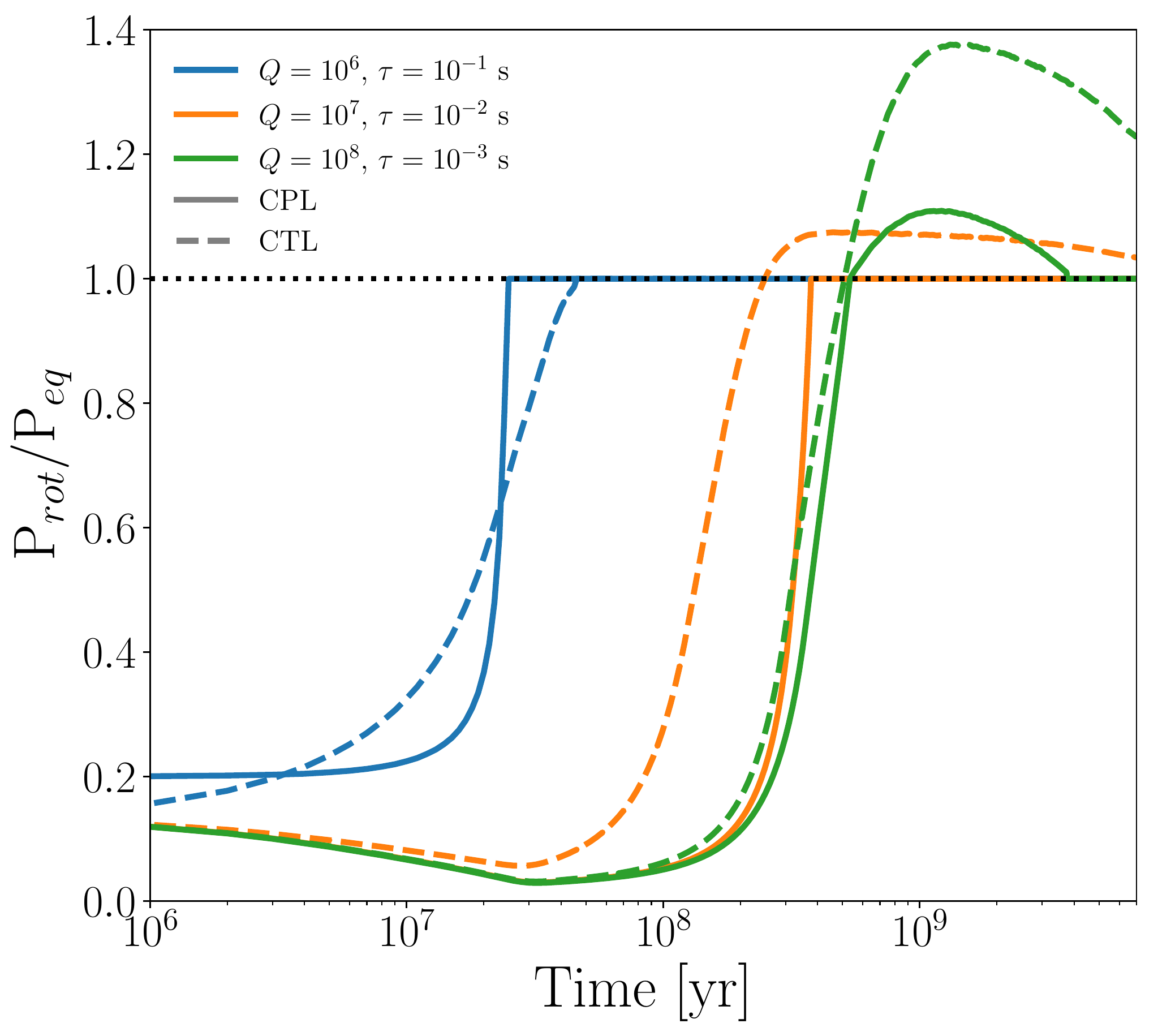}
   \caption{Evolution of stellar P$_{rot}$, normalized by P$_{eq}$ (see Eqn.~(\ref{eqn:cpl:eqPer}) and Eqn.~(\ref{eqn:ctl:eqPer})), for initial circular binary orbits with initial P$_{orb} = 7.5$ d according to the CPL (solid) and CTL (dashed) models for several values of $Q$ and $\tau$, respectively, using the \citet{Matt2015} magnetic braking model. Systems with strong tidal torques tidally lock, whereas in systems with weaker tidal torques (larger $Q$ and smaller $\tau$, respectively), magnetic braking initially overpowers tidal torques, spinning down the stars past the tidally locked state, resulting in subsynchronous rotation. }%
    \label{fig:eqPerShortPorb}%
\end{figure}

As seen in Fig.~\ref{fig:qTauLock}, subsynchronous rotation can even occur for short P$_{orb}$ binaries, where tidal-locking is the expectation, if the tidal torques are sufficiently weak. In Fig.~\ref{fig:eqPerShortPorb}, we examine subsynchronous rotation in short P$_{orb}$ binaries by displaying the P$_{rot}$ evolution for a P$_{orb} = 7.5$ d binary for various tidal dissipation parameters. Subsynchronous rotation occurs in general for weak tidal torques, $Q>10^7$ or $\tau < 0.1$ s in these cases, and is not restricted to long P$_{orb}$ binaries. Previous theoretical studies have also predicted subsynchronous rotation in short P$_{orb}$ binaries arising from the balance between tidal torques and magnetic braking \citep[e.g.][]{Habets1989,Zahn1994,Keppens1997} suggesting that this behavior is not an artifact of our choice of tidal or magnetic braking models, but rather a general outcome of the competition between magnetic braking and tidal evolution in low-mass binaries.  Short P$_{orb}$ subsynchronous rotators can eventually tidally lock after several Gyrs, e.g. the $Q=10^8$ case in Fig.~\ref{fig:eqPerShortPorb}, via the mechanism described above where tidal torques gradually strengthen relative to magnetic braking. 

Short P$_{orb}$ subsynchronous binaries exist in nature, such as many \kepler EBs (\citet{Lurie2017}, see $\S$~\ref{sec:kepler} for further discussion), Kepler-47 \citep{Orosz2012}, EPIC 219394517 \citep{Torres2018}, and in ``Binary 6211" observed by \citet{Meibom2006}, suggesting that this theoretical observation is real and borne out in nature. Spin-orbit synchronization should therefore not be assumed for short P$_{orb}$ binaries and sunsynchronous rotation should be expected in many tidally interacting binaries.  We explore these effect further and compare our theory to observations of \kepler EBs in $\S$~\ref{sec:kepler}.

\subsection{P$_{rot}$ Distribution of a Synthetic Population of Stellar Binaries} \label{sec:protDist}

\begin{figure}
	\includegraphics[width=0.5\textwidth]{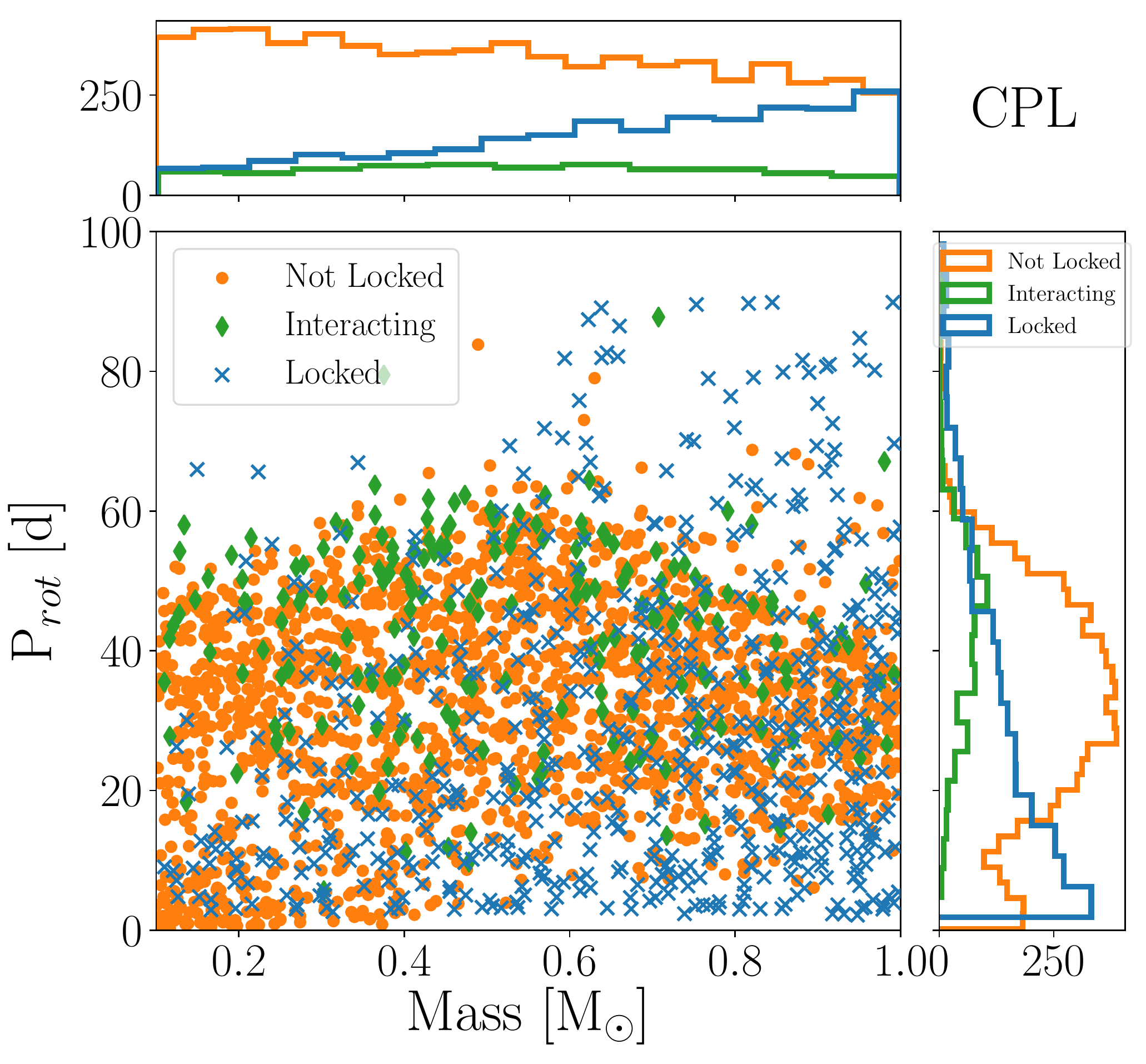}
   \caption{Rotation state for tidally locked (blue, P$_{rot} = $ P$_{eq}$), interacting (green, P$_{rot}$ within $10\%$ of P$_{eq}$ and not locked), and not locked (orange, remainder of binaries) stellar binaries. Left: P$_{rot}$ as a function of stellar mass and age according to our CPL simulations integrated to system ages uniformly sampled over $1-7$ Gyr. Right: Marginalized P$_{rot}$ distribution for each case. Top: Marginalized mass distributions.}%
    \label{fig:lockedCPL}%
\end{figure}

\begin{figure}
	\includegraphics[width=0.48\textwidth]{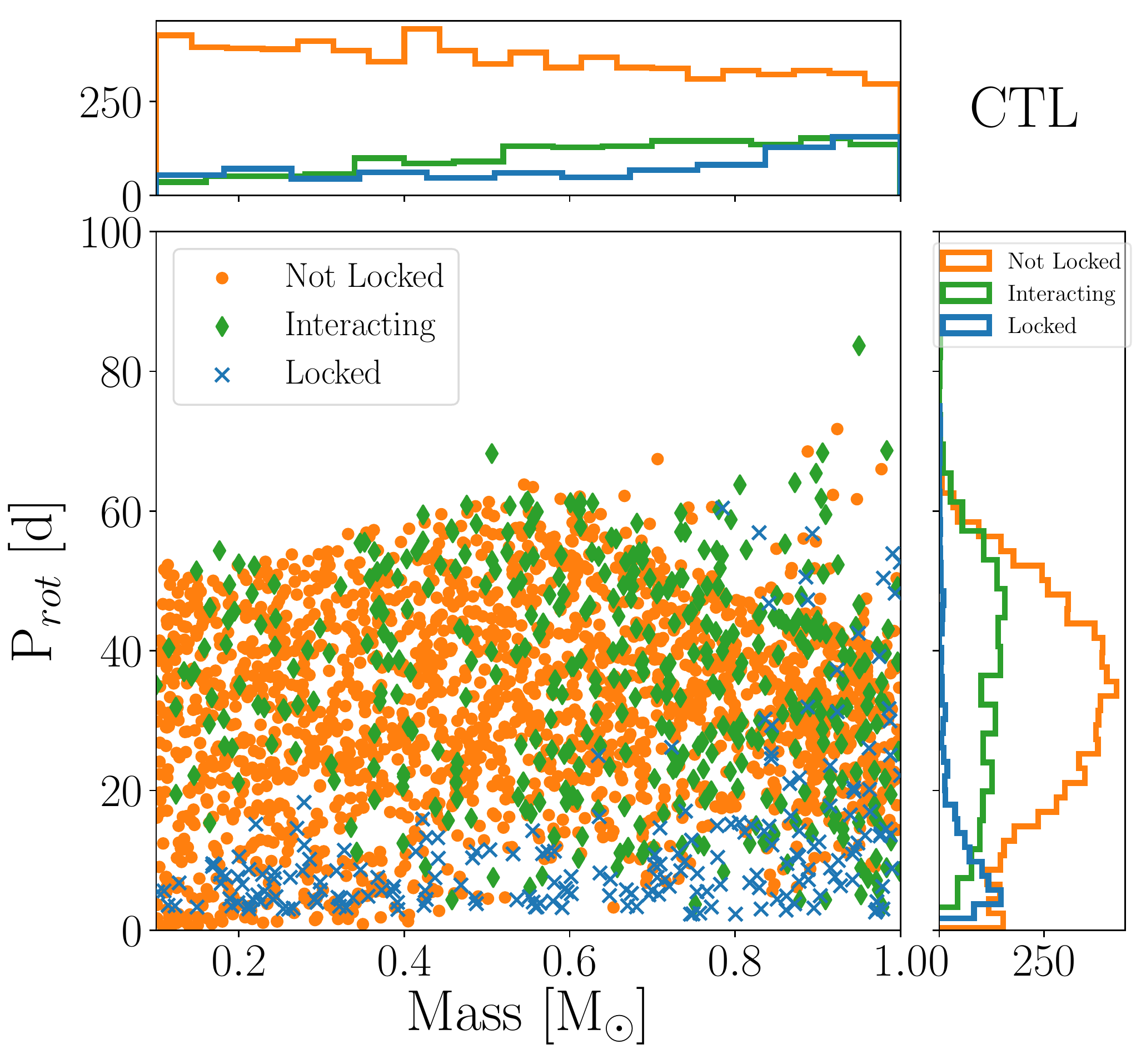}
   \caption{Same format as Fig.~\ref{fig:lockedCPL}, but for the CTL simulations.}%
    \label{fig:lockedCTL}%
\end{figure}

Here we examine how the competition between tidal torques and magnetic braking shape the P$_{rot}$ distribution of low-mass stellar binaries.  We consider two cases where tidal torques dominate: ``Locked", where P$_{rot} = $ P$_{eq}$, and ``Interacting", where P$_{rot}$ is within $10\%$ of P$_{eq}$ as in this regime, tides are likely shepherding P$_{rot}$ towards the tidally locked state as we demonstrated in $\S$~\ref{sec:eq}. We refer to the remaining binaries as ``not locked" as magnetic braking and stellar evolution likely dominate their angular momentum evolution.  In Fig.~\ref{fig:lockedCPL} and Fig.~\ref{fig:lockedCTL}, we plot P$_{rot}$ as a function of mass for the primary stars in stellar binaries for both the CPL and CTL model, respectively, integrated to system ages uniformly sampled over $1-7$ Gyr, consistent with ages of stars in the \kepler field \citep{Chaplin2014}.  

Both models predict a substantial population of tidally locked fast rotators with P$_{rot} \lsim 20$ d, with tidally locked stars systematically rotating faster (median CPL, CTL P$_{rot} = 22.6$ d and $8.8$ d) than not locked (median CPL, CTL both P$_{rot} = 32.4$ d) binaries. The CTL model predicts that the majority of tidally locked binaries, $83\%$, lock into rapid rotation with P$_{rot} \lsim 20$ d, typically in short P$_{orb}$ binaries where tidal torques are strongest. The CPL model, however predicts that binaries can tidally lock into a wide range of rotation states as only $46\%$ of locked binaries have P$_{rot} < 20$ d, while the rest can lock out to P$_{rot} \approx 100$ d in long P$_{orb}$ binaries. More massive stars are more likely to tidally lock compared to less massive stars as tidal torques scale with the stellar masses and as $R^5$, with $R$ increasing with stellar mass.  This feature is seen in the enhanced density of locked systems at larger masses for both tidal models, but in particular for the CPL model. We highlight this enhanced density of locked binaries in systems with more massive primaries in the marginalized mass distributions in the top panels of Fig.~\ref{fig:lockedCPL} and Fig.~\ref{fig:lockedCTL}.

The interacting population tends to rotate more slowly than the not locked population as at short P$_{orb}$, and hence P$_{rot}$, binaries preferentially tidally lock due to stronger tidal torques.  At longer P$_{orb}$, weaker tidal torques allow magnetic braking to spin down the stars past P$_{eq}$, with tidal torques eventually strengthening enough to shepherd P$_{rot}$ towards P$_{eq}$ via the mechanism discussed in $\S$~\ref{sec:eq}. The CPL and CTL models predict that $31\%$ and $24\%$ of stars, respectively, are either tidally locked or interacting, demonstrating that tidal torques play a pivotal role in shaping the angular momentum evolution in stellar binaries across a wide range of parameters. The P$_{rot}$ - mass distribution for not locked binaries resembles the single star sequence as magnetic braking and stellar evolution dictate their angular momentum evolution.

\begin{figure}
	\includegraphics[width=0.48\textwidth]{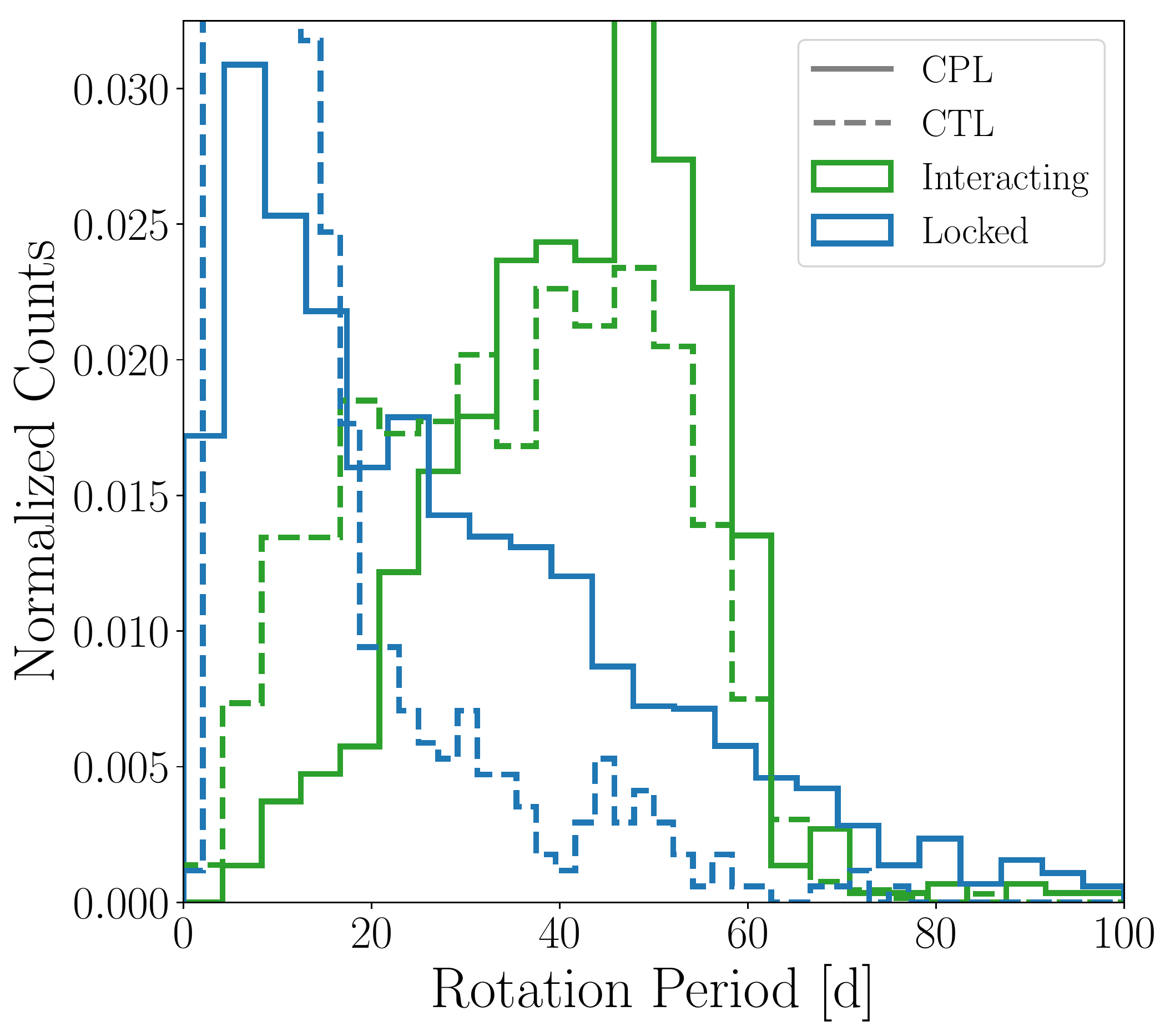}
   \caption{P$_{rot}$ distribution for tidally locked (blue) and interacting (green, P$_{rot}$ within $10\%$ of P$_{eq}$ and not locked) binaries according to the CPL (solid line) and CTL (dashed line) models.}%
    \label{fig:lockedProtHist}%
\end{figure}

Tidal locking is not limited to P$_{orb} \lsim 20$ d, however, as we find stellar binaries can tidally lock over a wide range of P$_{rot}$ up to P$_{rot} =$ P$_{orb} \approx 100$ d according to the CPL model, producing a slow-rotating population above the P$_{rot}$ distribution envelop of solar-mass single stars. This behavior is consistent with observations of P$_{rot}$ in \kepler eclipsing binaries by \citet{Lurie2017} who find tentative evidence that binaries can tidally lock up to their detection limit of P$_{orb} = $ P$_{rot} = 45$ d. Under the CTL model, however, binaries predominantly tidally lock out to only P$_{orb} \approx 20$ d, although binaries with more massive primaries can occasionally lock, or at least tidally-interact, out to P$_{orb} \approx 80$ d. We highlight this behavior with a histogram of locked and interacting binaries over P$_{rot}$ for both tidal models in Fig.~\ref{fig:lockedProtHist}. The CTL model predicts fewer tidally locked binaries at longer P$_{rot}$, concentrating most of the locked CTL distribution's density at short P$_{rot}$, whereas the CPL distribution has a heavy tail extending towards longer P$_{rot}$. The CPL model, however, predicts larger tidal-locking rates than the CTL model as seen in the enhanced numbers of tidally locked binaries at low P$_{rot}$ Fig.~\ref{fig:lockedCPL} compared with Fig.~\ref{fig:lockedCTL}. The presence of P$_{orb} > 20$ d locked population, or lack there of, could be a powerful observational discriminant between which equilibrium tidal model acts in low-mass stellar binaries. We discuss this point further in $\S$~\ref{sec:whichModel}. 

\subsection{Deviations From Single Star P$_{rot}$ Evolution: Implications for Gyrochronology} \label{sec:gyro}

\begin{figure*}[t]
	\includegraphics[width=\textwidth]{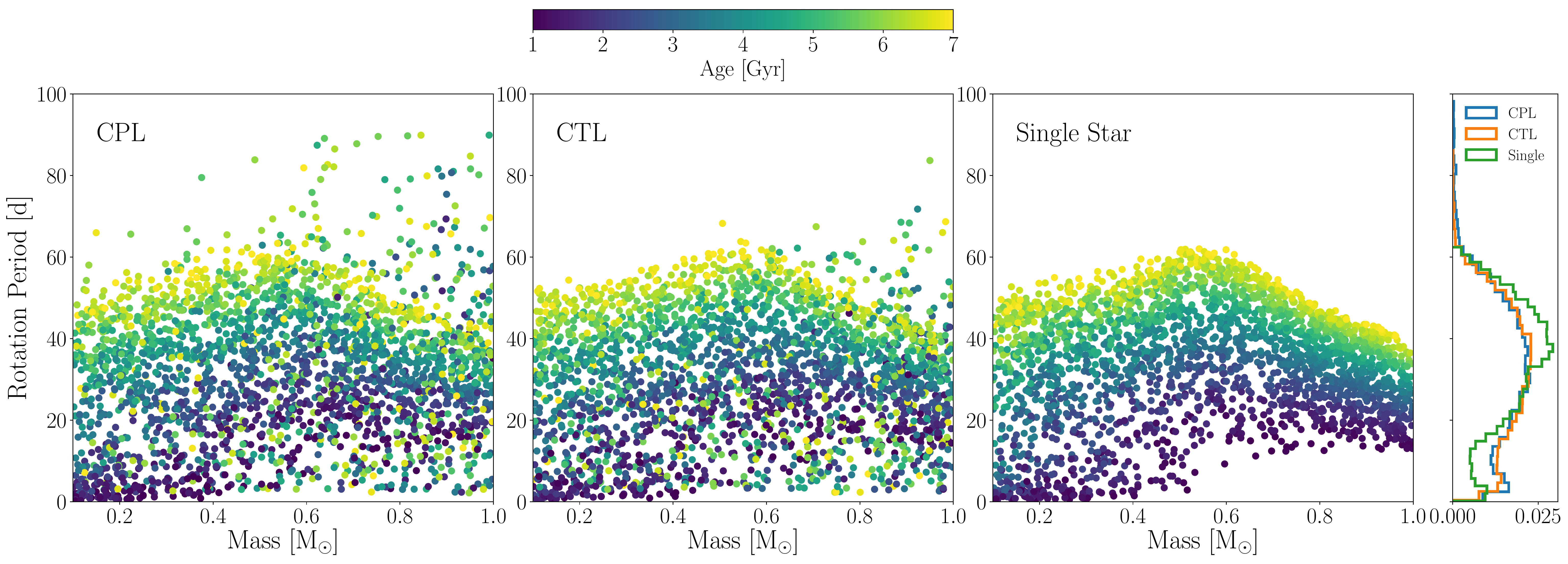}
   \caption{P$_{rot}$ as a function of stellar mass and age according to our CPL (left), CTL (left center), and single star (right center) simulations integrated to system ages uniformly sampled over $1-7$ Gyr using the \citet{Matt2015} magnetic braking model. For each case, we only plot 2,500 systems for clarity but account for all systems when computing the marginalized distributions. Right: The P$_{rot}$ distribution for each case, marginalized over stellar mass.}%
    \label{fig:protDist}%
\end{figure*}

We compare the P$_{rot}$ and age distributions of tidally interacting stellar binaries from our CPL and CTL simulations with that of single stars to gauge the impact of tidal torques on driving P$_{rot}$ distributions away from that of single stars and what implications that may have for estimating stellar ages using gyrochronology.  We simulate 10,000 single star systems according to the evolution described in $\S$~\ref{sec:methods:stellar} with initial conditions sampled from the same mass and P$_{rot}$ distributions used for the binary simulations described $\S$~\ref{sec:methods}. In Fig.~\ref{fig:protDist}, we display P$_{rot}$ as a function of mass and age for binaries simulated using both the CPL and CTL model and for single stars.

In binaries, tidal torques tend to drive the P$_{rot}$ evolution away from that of single stars and towards P$_{eq}$, either maintaining rapid rotation in tidally locked short P$_{orb}$ systems, or working with magnetic braking to slow P$_{rot}$ beyond that of single stars of the same age. The impact of tidal torques on the binary P$_{rot}$ distribution is clear: strong tidal torques in short P$_{orb}$ binaries produce a substantial population of rapid rotators with P$_{rot} \lsim 20$ d. Except for stars with ages $\lsim 1$ Gyr, or young late M-dwarfs who are either contracting along the pre-main sequence or have just reached the main sequence, our single star simulations fail to produce a population of rapid rotators. This theoretical result is consistent with \citet{Simonian2018} who find that the population of \kepler stars with P$_{rot} < 7.5$ d is likely dominated by tidally interacting binaries.  The influence of tides extends to longer P$_{orb}$ systems producing a slowly-rotating population above the upper envelope of the single star sequence for M$ \gsim 0.6$ M$_{\odot}$, e.g. Fig.~\ref{fig:lockedCPL} and Fig.~\ref{fig:lockedCTL}, and is a population that single-star models fail to produce. 

In the single star population, there is a clear monotonic relation between P$_{rot}$ and age, with older stars rotating more slowly, a trend that is borne out in nature and is the critical assumption of gyrochronology methods that link P$_{rot}$ to stellar ages via the magnetic braking-driven long-term spin down of low-mass stars \citep[e.g.][]{Skumanich1972,Barnes2003,Barnes2007,Mamajek2008,Barnes2010,Meibom2015}. This trend is a generic outcome of magnetic braking and is not specific to our choice of magnetic braking model. In stark contrast, both tidal models predict that age does not always strongly correlate with P$_{rot}$ as tidally interacting binaries, at a given primary star mass and P$_{rot}$, can assume a wide range of ages, especially for P$_{rot} \lsim 20$ d. 

\begin{figure*}[t]
	\includegraphics[width=\textwidth]{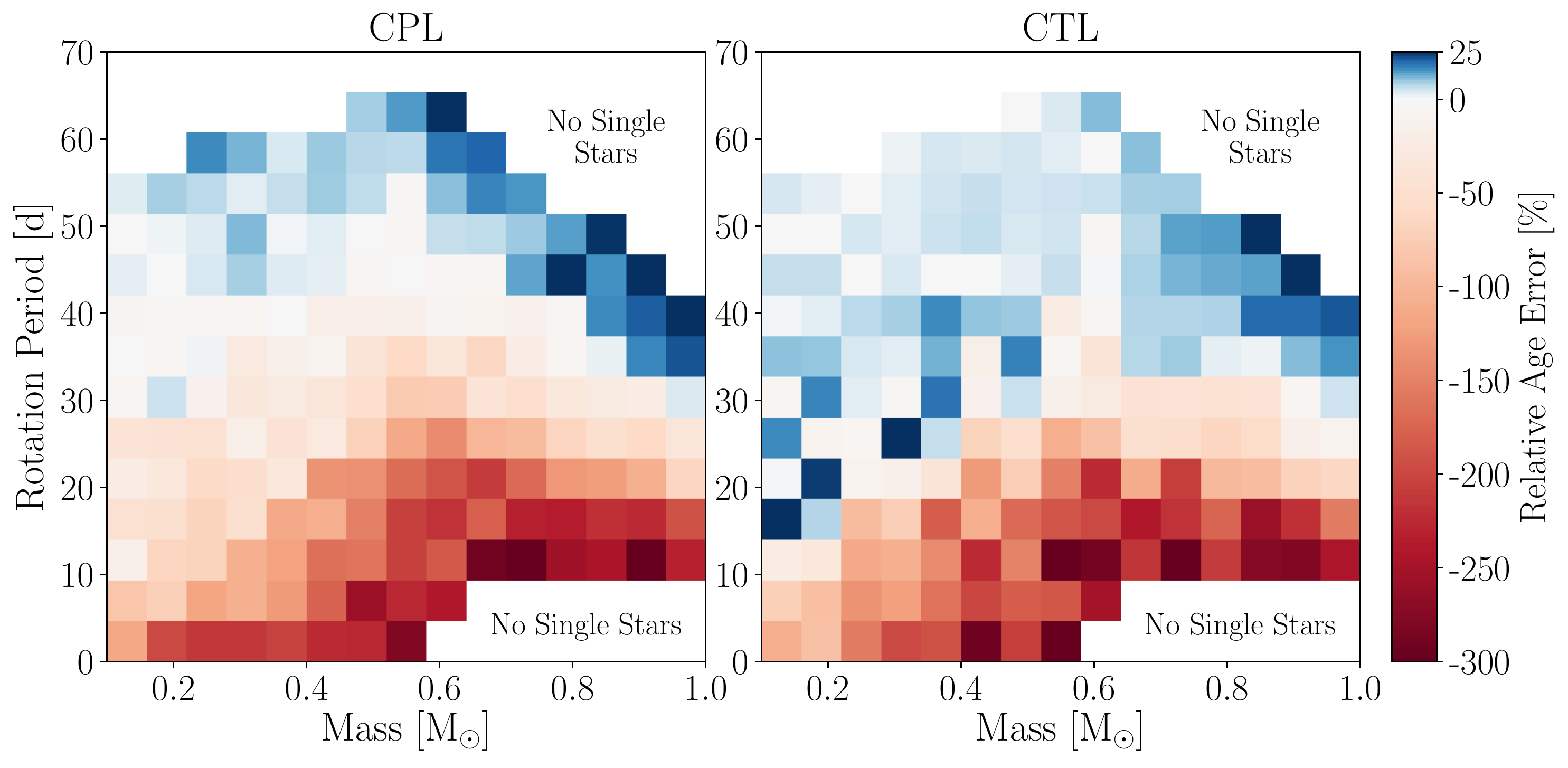}
   \caption{Relative age error between single and binary stars for both the CPL (left) and CTL (right) tidal models. The error is computed as the percent difference between the mean ages of single and tidally interacting binary stars in mass and P$_{rot}$ bins.}%
    \label{fig:gyro}%
\end{figure*}

We quantify the impact of binarity on gyrochronology age estimates in Fig.~\ref{fig:gyro} by computing the percent difference between the mean ages of single and binary stars, for both tidal models, in mass and P$_{rot}$ bins for the populations depicted in Fig.~\ref{fig:protDist}. This quantity, referred to here as the ``Relative Age Error", represents the systematic error incurred by assigning a tidally interacting binary star the age expected for single stars at a given mass and P$_{rot}$. For this comparison, we select the subset of tidally locked and tidally interacting binaries as classified in Fig.~\ref{fig:lockedCPL} and Fig.~\ref{fig:lockedCTL}. 

For most values of P$_{rot}$, gyrochronology methods systematically underestimate the ages of tidally interacting binaries, with the relative age error increasing with decreasing P$_{rot}$ and increasing with primary star mass. For binaries with P$_{rot} \lsim 30$ d, gyrochronology ages are underestimated by $50\%$, with this error growing to $300\%$ for P$_{rot} \approx 10$ d. For slow rotators with P$_{rot} \gsim 40$ d, gyrochronology ages are slightly overestimated by up to $25\%$, with the largest errors occurring for near solar-mass primary stars where tides and magnetic braking combine to spin down binary stars.  The relative age errors would be more pronounced for the most rapidly-rotating stars, e.g. tidally interacting near-solar mass binaries with P$_{rot} \lsim 10$ d, however, our single star-only models fail to produce such rotation states.

\begin{figure}[h]
	\includegraphics[width=0.45\textwidth]{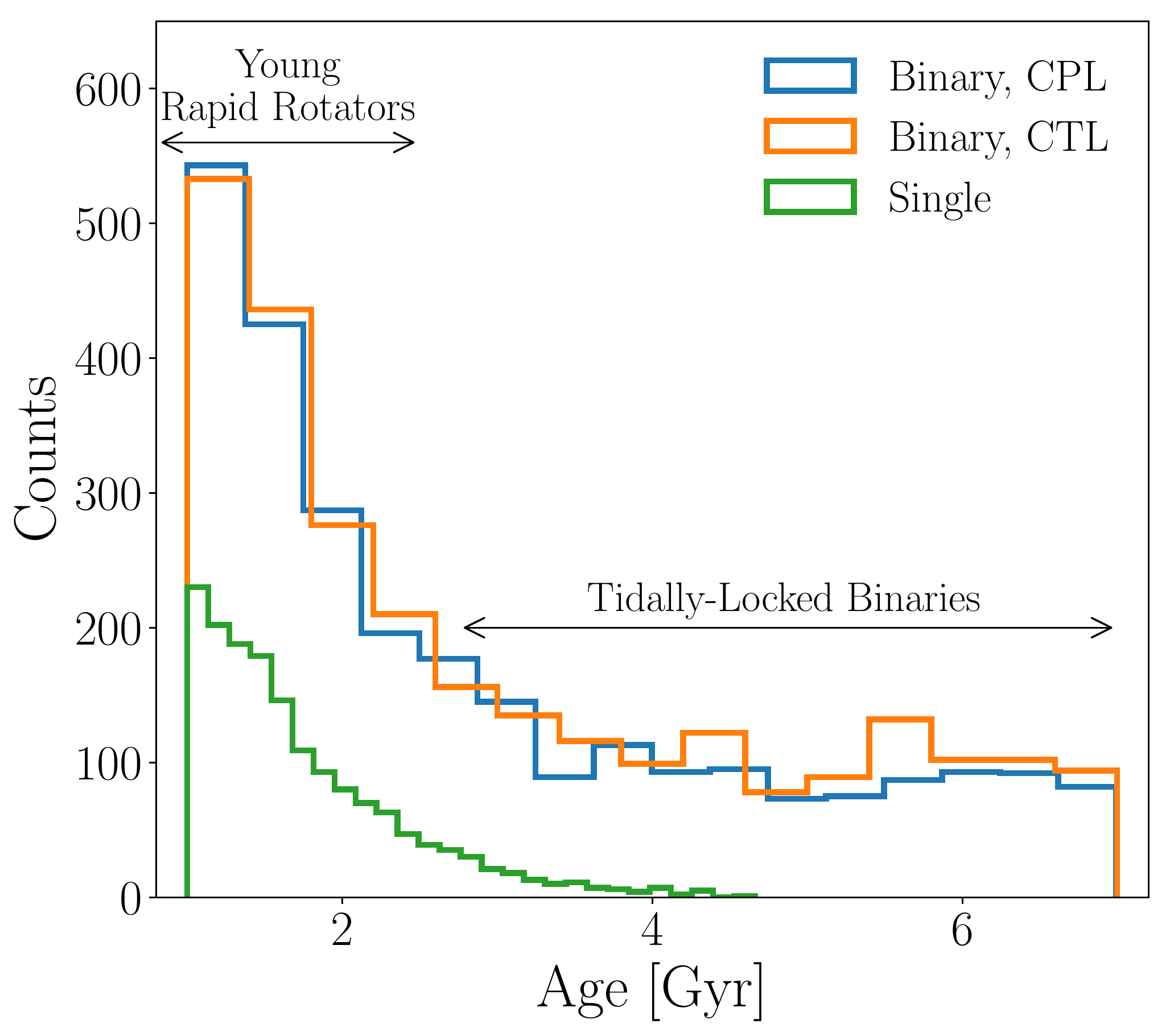}
   \caption{Histogram of rapidly-rotating (P$_{rot} < 20$ d) star ages for single and primary stars in binaries from Fig.~\ref{fig:protDist}. Rapidly-rotating single stars must be young (ages $\lsim 2$ Gyr), while tidally locked rapidly-rotating binaries exhibit a wide range of ages.}%
    \label{fig:protAgeHist}%
\end{figure}

The age distribution of binaries with P$_{rot} < 20$ d is inconsistent with that of single stars. In this range, the median ages and $68\%$ interval are $2.3^{+2.9}_{-0.9}$ Gyr and $2.4^{+3.0}_{-1.1}$ Gyr according to the CPL and CTL models, respectively, compared to the much younger single stars with ages of $1.6^{+0.8}_{-0.4}$ Gyr. We highlight this dichotomy in Fig.~\ref{fig:protAgeHist} by plotting a histogram of system ages from Fig.~\ref{fig:protDist} for single or primary stars in binaries with P$_{rot} < 20$ d. 

Tidal torques pose a fundamental problem for inferring ages of stars via gyrochronology. Regardless of the choice of equilibrium tidal model or magnetic braking model, stellar binaries readily tidally lock, or at least strongly tidally-interact, across a wide range of P$_{orb}$ and primary star masses, decoupling P$_{rot}$ from age. For example, if one observed a rapidly rotating star with P$_{rot} \lsim 20$ d, gyrochronology models would predict ages $\lsim 1.6$ Gyr. If the star is actually an unresolved binary, as could be the case for many \kepler rapid rotators \citep{Simonian2018}, it would likely be tidally locked, decoupling P$_{rot}$ from age, causing the predictions of gyrochronology models to fail. This effect is most likely to manifest in rapid rotators (P$_{rot} < 20$ d), but persists across all P$_{rot}$ up to 100 d, producing a contaminating signal, e.g. Fig.~\ref{fig:protDist} and Fig.~\ref{fig:gyro}.

In general, it is difficult to accurately determine if a source is single star or a stellar binary via longterm photometric monitoring, e.g. via \kepler or \textit{TESS}, as only a small fraction of stars in binaries will occult one another. Observations of the binarity of field stars by \citet{Raghavan2010} and \citet{Duchene2013} indicate that roughly half of stars are in stellar binaries, with $10\%$ of these binaries having P$_{orb} \lsim 100$ d, suggesting that unless one accounts for binarity, stellar binaries will produce a contaminating signal in any study of stellar rotation periods and any ages inferred via gyrochronology are potentially subject to systematic errors.  Moreover, this problem could be more significant as \citet{Simonian2018} found that most rapid rotators with P$_{rot} \leq 7.5$ d in the \kepler field are consistent with tidally-synchronized photometric binaries, suggesting that binary contamination in P$_{rot}$ studies could be widespread. We caution that any application of gyrochronology methods to predict ages for stars, especially those with P$_{rot} \lsim 20$ d, should rule out or account for stellar binarity, or otherwise risk deriving systematically incorrect ages. Tidal torques do not just produce spin-orbit synchronization at short P$_{orb}$, but can produce a rich variety of rotation states that deviate from the expected long-term spin-down experienced by single stars, e.g. Fig.~\ref{fig:qmap} and Fig.~\ref{fig:taumap}. We recommend that the application, or calibration, magnetic braking models to a sample of stellar rotation periods control for binarity.   

\subsection{Comparison to \kepler} \label{sec:kepler}

\begin{figure*}[t]
	\includegraphics[width=\textwidth]{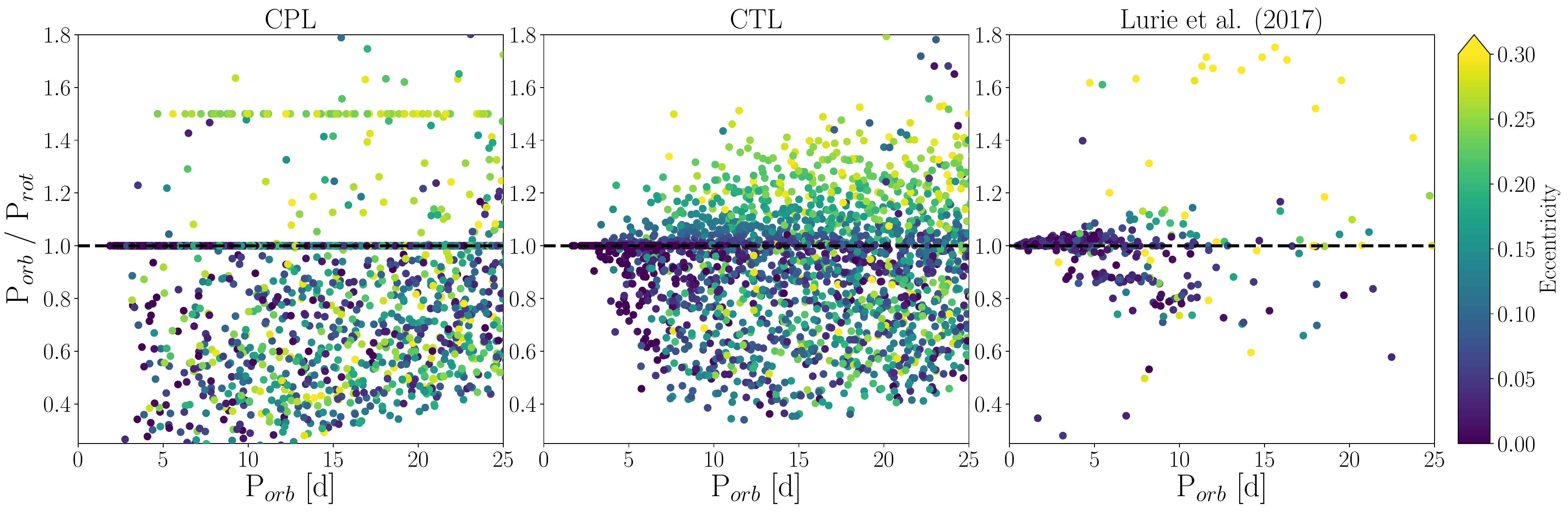}
   \caption{P$_{orb}/$P$_{rot}$ as a function of P$_{orb}$ according to the CPL model (left) and the CTL model (middle), and \citet{Lurie2017} \kepler EB observations (right). All points are colored by $e$.  In the right panel, the \kepler EBs at low P$_{orb}$ and low P$_{orb}/$P$_{rot}$ are likely either brown dwarfs or exoplanets \citep{Lurie2017}, and hence are not modeled by our simulations, so we do not consider them, but we display them for completeness.}%
    \label{fig:lurie7}%
\end{figure*}

We compare our simulation results to P$_{rot}$ measurements of primary stars in \kepler low-mass eclipsing binaries by \citet{Lurie2017} to gauge if our model predictions, which by design populate a wide, but physically-plausible, region of parameter space, can reproduce features observed in the data.  \citet{Lurie2017} measured 816 rotation periods for primary stars in \kepler EBs with star spot modulations and visually inspected each light curve to ensure their accuracy. The \citet{Lurie2017} dataset is the largest homogenous set of P$_{rot}$ measurements available for low-mass stellar binaries and represents the state of the art benchmark for studies of the influence of tides on P$_{rot}$ in stellar binaries. We compare our results to the P$_{1,min}$ P$_{rot}$ values reported by \citet{Lurie2017} as the authors demonstrated that these values are likely to be close to the equatorial P$_{rot}$ that we track in our simulations. In Fig.~\ref{fig:lurie7}, we display P$_{orb}/$P$_{rot}$ as a function of P$_{orb}$ for both the CPL and CTL models where each simulation was integrated to an age uniformly sampled over $1-7$ Gyr, consistent with ages of \kepler field stars \citep{Chaplin2014}. 

Qualitatively, the CTL model appears to do a better job of reproducing features seen in the \citet{Lurie2017} data than the CPL model. The CPL model, for example, cannot produce the observed cluster of supersynchronous rotators with P$_{orb}/$P$_{rot} \lsim 1.2$ for P$_{orb} < 10$ d whereas the CTL model can. Instead, owing to the its discrete P$_{eq}$, the CPL model predicts that all tidally locked supersynchronous rotators lie on the line P$_{orb}/$P$_{rot} = 1.5$. This prediction is inconsistent with the data as no obvious spin-orbit commensurablity, aside from 1:1 synchronization, is present in the \citet{Lurie2017} data, likely because stellar convective envelopes lack a fixed shape, making resonant coupling difficult unless it occurs with internal gravity or pressure modes \citep{Burkart2014,Lurie2017}. Neither model reproduces the very supersynchronous, P$_{orb}/$P$_{rot} > 1.6$, binaries in the \citet{Lurie2017} data as they typically have orbital eccentricities in excess of 0.3 and are outside of the region of parameter space we consider. The CTL model, however, could in principle reproduce these points if they are tidally locked binaries as its P$_{eq}$ is a continuous function of $e$ and is applicable for large $e$, in contrast to the discrete P$_{eq}$ predicted by the CPL model that is valid for smaller $e$. Both tidal models predict that nearly all binaries with P$_{orb} < 4$ d have circularized orbits and synchronized spins due to strong tidal torques at short stellar separations, in agreement with the \citet{Lurie2017} observations. At very short P$_{orb}$, in the absence of a perturbing tertiary companion, circularization and synchronization is the inevitable end state for low-mass binaries \citep{Counselman1973}. 

For P$_{orb} \gsim 4$ d, our models produce a substantial number of subsynchronous rotators. Although \citet{Lurie2017} argues that differential rotation creates the subsynchronous population, we find that the competition between weak tidal torques and magnetic braking described in $\S$~\ref{sec:eq} naturally produces this population. The CPL model, however, struggles to populate the prominent cluster of subsynchronous rotators at P$_{orb}/$P$_{rot} \approx 0.9$ for P$_{orb} < 10$ d observed by \citet{Lurie2017}.  \citet{Lurie2017} find that $15\%$ of their sample with 2 $< P_{orb} <$ 10 days has P$_{orb}/$P$_{rot} \in [0.84, 0.92]$, compared with $8\%$ of our CTL population and only $2\%$ of the CPL population.

Both models predict a large number of extremely subsynchronous rotators with P$_{orb}/$P$_{rot} < 0.7$ across all P$_{orb}$ that is not present in the \citet{Lurie2017} data. Magnetic braking creates the lower limit of this subsynchronous population, a line of nearly constant P$_{rot} \approx 60$ d set by how much a star can spin down over 7 Gyr, the longest age considered in our simulations. Our choice of prior distributions for both $Q$ and $\tau$ permit very weak tidal interactions that likely gives rise to this population and suggests that our prior does not reflect the underlying distribution of stellar tidal parameters in nature. Alternatively, the data could be incomplete where our models predict slowly-rotating subsynchronous rotators as the photometric amplitude of star spot modulations tends to decrease with increasing P$_{rot}$, making reliable rotation periods difficult to detect \citep{McQuillan2014,Lurie2017,Reinhold2018}. 

Although the CTL model seems to better reproduce the \citet{Lurie2017} data, both tidal models can reproduce features observed in the \kepler EB distribution, e.g. the synchronized population and subsynchronous rotators, suggesting that our models reasonably approximate the dynamical interactions of tidally-evolving, low-mass stellar binaries. Our comparison between theory and observations is limited, however, because the \citet{Lurie2017} P$_{rot}$ data lack uncertainties and \citet{Lurie2017} approximated the EB orbital $e$ via transit durations and ingress/egress times, potentially leading to inaccurate $e$ determinations. Unconstrained biases in the data, e.g. the lack of long P$_{orb}$ binaries, further inhibit our ability to compare our predictions with the data. Moreover, our prior distributions were chosen to be plausible, but wide, in order to examine our model predictions over parameter space and are not suited for a robust statistical inference to select between which equilibrium tidal model best describes tidal interactions in low-mass binaries stars. Below, we offer observational tests that could discriminate between models.

\subsection{CPL or CTL?} \label{sec:whichModel}

Accurate measurements of P$_{rot}$ and $e$, especially out to long P$_{orb}$, can potentially discriminate between which equilibrium tidal model best describes tidal interactions in low-mass stellar binaries. Here, we outline three observational tests that can discriminate between the two models. The first test considers binaries with P$_{orb} < 10$ d that are likely tidally locked on eccentric orbits, but with $e < 0.23$.  In this $e$ regime, the CPL model predicts that the majority of systems are tidally locked into synchronous rotation and does not permit a supersynchronous rotation state, e.g. Eqn.~(\ref{eqn:cpl:eqPer}). The CTL model, however, predicts a continuum of supersynchronous rotators on eccentric orbits, e.g. Eqn.~(\ref{eqn:ctl:eqPer}). Supersynchronous rotation that is not due to tidal interactions can occur in extremely young, rapidly rotating systems that are still contracting along the pre-main sequence, or that have recently reached the main sequence.  These young, supersynchronous rotators are unlikely to be tidally locked, usually have P$_{orb}/$P$_{rot} > 1.5$, and do not stay supersynchronous for long given that solar mass pre-main sequence lifetimes are $\lsim 100$ Myr, distinguishing them from tidally locked binaries (see Fig.~\ref{fig:lurie7}).   If supersynchronous rotation is observed in binaries with P$_{orb} < 10$ d, P$_{orb}/$P$_{rot} < 1.5$, and $0 < e \lsim 0.23$, it is evidence in favor of the CTL model over the CPL model. 

Second, for tidally locked binaries with $e > 0.23$, the CPL model predicts supersynchronous rotation in the form of a 3:2 spin-orbit comensurability, e.g. the line at P$_{orb}/$P$_{rot} = 1.5$ seen in the left panel of Fig.~\ref{fig:lurie7}, and no other spin state is permitted, compared to the continuum of supersynchronous rotation states in eccentric tidally locked rotators predicted by the CTL model. If a substantial clustering of stellar binaries with P$_{orb}/$P$_{rot} = 1.5$ is observed, it would be strong evidence in favor of the CPL model, but there is no obvious clustering of \kepler EBs near any spin-orbit resonance. These two tests can fail to discriminate between the CPL and CTL model, however, if the CPL model P$_{eq}$ is a continuous function of $e$, e.g. Eqn.~(\ref{eqn:cpl:eqPerCont}), as was argued by \citet{Goldreich1966b} and derived by \citet{Murray1999}.  In such a case, one would need a large number of accurate and precise measurements P$_{orb}$ and $e$, with robust uncertainties, for tidally interacting binaries to discriminate between the CPL and CTL continuous P$_{eq}$, e.g. Eqn.~(\ref{eqn:cpl:eqPerCont}) versus Eqn.~(\ref{eqn:ctl:eqPer}).  In practice, this is extremely observationally expensive as it requires extensive photometric and spectroscopic observations of many binaries. 

 A third test, the detection of tidally locked binaries with solar-mass primaries and P$_{rot} \gsim 60$ d, would provide strong evidence in favor of the CPL model as the CTL model cannot tidally lock stars beyond P$_{orb} \approx 60$ d, regardless of $\tau$, e.g. Fig.~\ref{fig:lockedCTL}. The CPL model, however, can tidally lock binaries out to P$_{orb} \gsim 90$ d.  We recommend observers try to measure P$_{rot}$ and $e$ in binaries out to P$_{orb} = 100 $ d to test this hypothesis, but we note that detecting P$_{rot}$ for such slow rotators can be difficult due to small star spot modulation amplitudes \citep{McQuillan2014,Lurie2017,Reinhold2018}. Long term spectroscopic monitoring may be warranted in such cases.

\section{Discussion} \label{sec:discussion}

In this work, we probed the long-term angular momentum evolution of low-mass stellar binaries, with a focus on P$_{rot}$ in short and intermediate P$_{orb}$ binaries.  We considered the impact of two common equilibrium tidal models, magnetic braking, and stellar evolution.  We performed a large suite of simulations for binaries with physically-motivated initial conditions out to P$_{orb} = 100$ and across a wide range of tidal dissipation parameters to examine the competition between tidal torques and magnetic braking for controlling the stellar P$_{rot}$ evolution. 

In our simulations, nearly all binaries with $P_{orb} \lsim 4$ d have tidally-synchronized spins and circularized orbits, in good agreement with observations of \kepler EBs and binaries in the field. We showed for P$_{orb} \gsim 4$ d, primary stars in stellar binaries can rotate subsynchronously for Gyrs due to the competition between tidal torques and magnetic braking, or supersynchronously if they tidally lock on eccentric orbits. Our predictions are not strongly dependant on the choice of magnetic braking model, but rather are generic outcomes of the interaction between magnetic braking and tidal torques.  Both the CPL and CTL equilibrium tidal models predict that binaries tidally-interact at longer P$_{orb}$ than have previously been considered, out to P$_{orb} \approx 60-100$ d. Many binaries with P$_{orb} \lsim 20$ d tidally lock according to both models, in good agreement with previous results, but the CPL model predicts that binaries can readily tidally lock out to P$_{orb} \approx 100$ d. Tidal interactions can cause P$_{rot}$ evolution in stellar binaries to differ from the long-term spin down due to magnetic braking experienced by single stars, decoupling P$_{rot}$ from age.  In tidally interacting binaries, gyrochronology, the technique of linking stellar P$_{rot}$ to age, likely fails, potentially underestimating stellar ages by up to $300\%$. We caution that any application of gyrochronology methods to stars, especially those with P$_{rot} \lsim 20$ d, should account for the possibility of stellar binarity to prevent deriving incorrect ages. 

We compare the predictions of both the CPL and CTL models with observations of P$_{rot}$ and P$_{orb}$ of \kepler EBs by \citet{Lurie2017} and find that both can qualitatively reproduce many features seen in the data, validating our approach and suggesting that equilibrium tidal models can accurately model stellar-tidal evolution in low-mass stellar binaries. The lack of uncertainties on P$_{rot}$, the approximate orbital eccentricities derived by \citet{Lurie2017}, and unconstrained completeness estimates prevent us from discriminating between which tidal model best describes tidal torques in low-mass binaries and from inferring tidal properties of low-mass stars given the \kepler EB data.  

We described three observational tests that can distinguish between which equilibrium tidal model better describes tidal interactions in low-mass stellar binaries. We primarily suggest that observers measure stellar P$_{rot}$ in binaries with solar-mass primaries for P$_{orb}$ between $60-100$ d. If any tidally locked binaries are identified at long P$_{orb}$, this would be evidence in favor of the CPL model as we found that only binaries tidally interacting via the CPL model could tidally lock at such long P$_{orb}$. At shorter P$_{orb}$, precise measurements of P$_{rot}$ and binary $e$ and P$_{orb}$ could distinguish between the CPL and CTL model in tidally locked systems, e.g. identifying if P$_{eq}$ follows Eq.~(\ref{eqn:cpl:eqPer}) vs. Eq.~(\ref{eqn:ctl:eqPer}), especially if the CPL P$_{eq}$ is in fact a discrete function of $e$. The observations required by these tests, however, are non-trivial. Beyond these tests, our model could be used to infer the tidal properties of binary stars, perhaps in a Markov Chain Monte Carlo framework, by directly comparing simulation results with the observed stellar and orbital properties, given the observational uncertainties and reasonable prior probability distributions for parameters like the initial binary $e$. This analysis, however, is beyond the scope of this work and we leave it for future endeavors.

Our theoretical predictions outline a critical point: one cannot simply observe a short P$_{orb}$ binary on a circular orbit and assume synchronization, nor can one observe a binary with P$_{orb} \gsim 20$ d and assume that tides have not impacted that system's angular momentum evolution.  Stellar-tidal interactions can produce synchronous and subsynchronous rotation for short P$_{orb}$ binaries on circular orbits, e.g. Fig.~\ref{fig:eqPer}, depending on the age of the system, e.g. Fig.~\ref{fig:eqPerShortPorb}, and the strength of tidal dissipation, e.g. Fig.~\ref{fig:qmap} and Fig.~\ref{fig:taumap}.  Understanding the long-term angular momentum evolution of stellar binaries out to P$_{orb} = 100$ d requires detailed modeling of its coupled-stellar tidal evolution, and characterizing tidal dissipation parameters. Many new eclipsing stellar binaries will be discovered by TESS \citep[e.g.][]{Sullivan2015,Matson2018} and in analysis of K2 data.  Obtaining precise orbital and rotational constraints for stellar binaries will permit detailed characterization of tidal interactions between low-mass stars and shed light into the long-term angular momentum evolution in stellar binaries. 

\acknowledgments
We thank the anonymous reviewer for helpful comments that improved the quality of this manuscript. This work was facilitated though the use of advanced computational, storage, and networking infrastructure provided by the Hyak supercomputer system and funded by the Student Technology Fund at the University of Washington. DPF was supported by NASA Headquarters under the NASA Earth and Space Science Fellowship Program - Grant 80NSSC17K0482.  RB acknowledges support from the NASA Astrobiology Institute's Virtual Planetary Laboratory under Cooperative Agreement number NNA13AA93A. JRAD acknowledges support from the DIRAC Institute in the Department of Astronomy at the University of Washington. The DIRAC Institute is supported through generous gifts from the Charles and Lisa Simonyi Fund for Arts and Sciences, and the Washington Research Foundation.

\software{matplotlib: \citet{Hunter2007}, numpy: \citet{vanderWalt2011}, pandas: \citet{Mckinney2010}, \vplanet: \citet{Barnes2019}}

\appendix

\section{Analytic Torque Balance} \label{sec:appendix:balance}

Here we derive the equation for the stellar P$_{rot}$ at which tidal torques balance magnetic braking discussed in $\S$~\ref{sec:analytic}. As in $\S$~\ref{sec:analytic}, we assume that both stars have M $= 1$M$_{\odot}$, 0 obliquity, and we assume a circular binary orbit. We assume that the torque balance occurs while the stars are on the main sequence, where stellar properties change slowly, so the angular momentum evolution is controlled by the balance between tidal torques and magnetic braking, not stellar radius contraction. Under this assumption, we can set $R = 1 R_{\odot}$ and assume constant moments of inertia. For simplicity, we assume that magnetic braking proceeds under the \citet{Matt2015} model and the CTL model describes tidal torques.

As discussed in $\S$~\ref{sec:analytic}, both stars are in the unsaturated rotation regime, so the torque due to magnetic braking is given by Eqn.~(\ref{eqn:mattUnSat}), which under the aforementioned assumptions, reduces to
\begin{equation}
    \frac{dJ}{dt}\Bigg|_{MB} = -C_{MB} \left( \frac{P_{rot,\odot}}{P_{rot}} \right)^3
\end{equation}
where P$_{rot} = 2 \pi / \omega$ and $C_{MB} = 6.3 \times 10^{30}$ ergs \citep{Matt2015,Matt2019}.

Under the CTL model and our assumptions, the change in rotation rate due to tidal torques, Eqn.~\ref{eqn:ctl:omega}, reduces to 
\begin{equation}
    \frac{d\omega}{dt}\Bigg|_{tides} = \frac{\mathrm{P}_{orb}\mathrm{Z}_{CTL}}{2 \pi M r_g^2 R^2}\left( 1 - \frac{\mathrm{P}_{orb}}{\mathrm{P}_{rot}} \right)
\end{equation}
where P$_{orb} = 2 \pi/n$.  For fixed moment of inertia, $dJ/dt =I d\omega/dt$, and after inserting Eqn.~\ref{eqn:ctl:z} for Z$_{CTL}$, the tidal torque on the stellar rotations becomes
\begin{equation}
    \frac{dJ}{dt}\Bigg|_{tides} = \frac{C_{tides} k_2 \tau}{\mathrm{P}_{orb}^5}\left( 1 - \frac{\mathrm{P}_{orb}}{\mathrm{P}_{rot}} \right).
\end{equation}
where $C_{tides} = 24 \pi^5 R_{\odot}^5 / G$. 

The torques due to tides and magnetic braking balance when $ \frac{dJ}{dt}|_{tides} +  \frac{dJ}{dt}|_{MB} = 0$, 
\begin{equation} \label{eqn:appendixBalance}
    \frac{C_{tides} k_2 \tau}{\mathrm{P}_{orb}^5}\left( 1 - \frac{\mathrm{P}_{orb}}{\mathrm{P}_{rot}} \right) -C_{MB} \left( \frac{P_{rot,\odot}}{P_{rot}} \right)^3 = 0.
\end{equation}
By specifying P$_{orb}$ and $k_2 \tau$, we can numerically solve Eqn.~(\ref{eqn:appendixBalance}) for the P$_{rot}$ at which torques due to magnetic braking and tides balance, often producing subsynchronous rotation as seen in Fig.~\ref{fig:analyticBalance} and our simulations in $\S$~\ref{sec:eq}.

\bibliography{sync}

\begin{thebibliography}{}
\expandafter\ifx\csname natexlab\endcsname\relax\def\natexlab#1{#1}\fi
\providecommand{\url}[1]{\href{#1}{#1}}

\bibitem[{{Ag{\"u}eros} {et~al.}(2011){Ag{\"u}eros}, {Covey}, {Lemonias},
  {Law}, {Kraus}, {Batalha}, {Bloom}, {Cenko}, {Kasliwal}, {Kulkarni},
  {Nugent}, {Ofek}, {Poznanski}, \& {Quimby}}]{Agueros2011}
{Ag{\"u}eros}, M.~A., {Covey}, K.~R., {Lemonias}, J.~J., {et~al.} 2011, \apj,
  740, 110

\bibitem[{{Allain}(1998)}]{Allain1998}
{Allain}, S. 1998, \aap, 333, 629

\bibitem[{{Baraffe} {et~al.}(2015){Baraffe}, {Homeier}, {Allard}, \&
  {Chabrier}}]{Baraffe2015}
{Baraffe}, I., {Homeier}, D., {Allard}, F., \& {Chabrier}, G. 2015, \aap, 577,
  A42

\bibitem[{{Barker} \& {Ogilvie}(2009)}]{Barker2009}
{Barker}, A.~J., \& {Ogilvie}, G.~I. 2009, \mnras, 395, 2268

\bibitem[{{Barnes}(2017)}]{Barnes2017}
{Barnes}, R. 2017, ArXiv e-prints, arXiv:1708.02981

\bibitem[{{Barnes} {et~al.}(2013){Barnes}, {Mullins}, {Goldblatt}, {Meadows},
  {Kasting}, \& {Heller}}]{Barnes2013}
{Barnes}, R., {Mullins}, K., {Goldblatt}, C., {et~al.} 2013, Astrobiology, 13,
  225

\bibitem[{{Barnes} {et~al.}(2019){Barnes}, {Luger}, {Deitrick}, {Driscoll},
  {Quinn}, {Fleming}, {Smotherman}, {McDonald}, {Wilhelm}, {Garcia}, {Barth},
  {Guyer}, {Meadows}, {Bitz}, {Gupta}, {Domagal-Goldman}, \&
  {Armstrong}}]{Barnes2019}
{Barnes}, R., {Luger}, R., {Deitrick}, R., {et~al.} 2019, arXiv e-prints,
  arXiv:1905.06367

\bibitem[{{Barnes}(2003)}]{Barnes2003}
{Barnes}, S.~A. 2003, \apj, 586, 464

\bibitem[{{Barnes}(2007)}]{Barnes2007}
---. 2007, \apj, 669, 1167

\bibitem[{{Barnes}(2010)}]{Barnes2010}
---. 2010, \apj, 722, 222

\bibitem[{{Bate}(2000)}]{Bate2000}
{Bate}, M.~R. 2000, \mnras, 314, 33

\bibitem[{{Bate} {et~al.}(2002){Bate}, {Bonnell}, \& {Bromm}}]{Bate2002}
{Bate}, M.~R., {Bonnell}, I.~A., \& {Bromm}, V. 2002, \mnras, 336, 705

\bibitem[{{Bolmont} \& {Mathis}(2016)}]{Bolmont2016}
{Bolmont}, E., \& {Mathis}, S. 2016, Celestial Mechanics and Dynamical
  Astronomy, 126, 275

\bibitem[{{Bonnell} \& {Bate}(1994)}]{Bonnell1994}
{Bonnell}, I.~A., \& {Bate}, M.~R. 1994, \mnras, 271, astro-ph/9411081

\bibitem[{{Bouvier}(2008)}]{Bouvier2008}
{Bouvier}, J. 2008, \aap, 489, L53

\bibitem[{{Burkart} {et~al.}(2014){Burkart}, {Quataert}, \&
  {Arras}}]{Burkart2014}
{Burkart}, J., {Quataert}, E., \& {Arras}, P. 2014, \mnras, 443, 2957

\bibitem[{{Chaplin} {et~al.}(2014){Chaplin}, {Basu}, {Huber}, {Serenelli},
  {Casagrande}, {Silva Aguirre}, {Ball}, {Creevey}, {Gizon}, {Handberg},
  {Karoff}, {Lutz}, {Marques}, {Miglio}, {Stello}, {Suran}, {Pricopi},
  {Metcalfe}, {Monteiro}, {Molenda-{\.Z}akowicz}, {Appourchaux},
  {Christensen-Dalsgaard}, {Elsworth}, {Garc{\'{\i}}a}, {Houdek}, {Kjeldsen},
  {Bonanno}, {Campante}, {Corsaro}, {Gaulme}, {Hekker}, {Mathur}, {Mosser},
  {R{\'e}gulo}, \& {Salabert}}]{Chaplin2014}
{Chaplin}, W.~J., {Basu}, S., {Huber}, D., {et~al.} 2014, \apjs, 210, 1

\bibitem[{{Claret} {et~al.}(1995){Claret}, {Gimenez}, \& {Cunha}}]{Claret1995}
{Claret}, A., {Gimenez}, A., \& {Cunha}, N.~C.~S. 1995, \aap, 299, 724

\bibitem[{{Counselman}(1973)}]{Counselman1973}
{Counselman}, III, C.~C. 1973, \apj, 180, 307

\bibitem[{{Cranmer} \& {Saar}(2011)}]{Cranmer2011}
{Cranmer}, S.~R., \& {Saar}, S.~H. 2011, \apj, 741, 54

\bibitem[{{Darwin}(1880)}]{Darwin1880}
{Darwin}, G.~H. 1880, Philosophical Transactions of the Royal Society of London
  Series I, 171, 713

\bibitem[{{Douglas} {et~al.}(2017){Douglas}, {Ag{\"u}eros}, {Covey}, \&
  {Kraus}}]{Douglas2017}
{Douglas}, S.~T., {Ag{\"u}eros}, M.~A., {Covey}, K.~R., \& {Kraus}, A. 2017,
  \apj, 842, 83

\bibitem[{{Duch{\^e}ne} \& {Kraus}(2013)}]{Duchene2013}
{Duch{\^e}ne}, G., \& {Kraus}, A. 2013, \araa, 51, 269

\bibitem[{{Fabrycky} \& {Tremaine}(2007)}]{Fabrycky2007}
{Fabrycky}, D., \& {Tremaine}, S. 2007, \apj, 669, 1298

\bibitem[{{Ferraz-Mello} {et~al.}(2008){Ferraz-Mello}, {Rodr{\'{\i}}guez}, \&
  {Hussmann}}]{FerrazMello2008}
{Ferraz-Mello}, S., {Rodr{\'{\i}}guez}, A., \& {Hussmann}, H. 2008, Celestial
  Mechanics and Dynamical Astronomy, 101, 171

\bibitem[{{Fleming} {et~al.}(2018){Fleming}, {Barnes}, {Graham}, {Luger}, \&
  {Quinn}}]{Fleming2018}
{Fleming}, D.~P., {Barnes}, R., {Graham}, D.~E., {Luger}, R., \& {Quinn}, T.~R.
  2018, \apj, 858, 86

\bibitem[{{Fleming} \& {Quinn}(2017)}]{Fleming2017}
{Fleming}, D.~P., \& {Quinn}, T.~R. 2017, \mnras, 464, 3343

\bibitem[{{Gaia Collaboration} {et~al.}(2016){Gaia Collaboration}, {Prusti},
  {de Bruijne}, {Brown}, {Vallenari}, {Babusiaux}, {Bailer-Jones}, {Bastian},
  {Biermann}, {Evans}, \& et~al.}]{Gaia2016}
{Gaia Collaboration}, {Prusti}, T., {de Bruijne}, J.~H.~J., {et~al.} 2016,
  \aap, 595, A1

\bibitem[{{Gallet} {et~al.}(2017){Gallet}, {Bolmont}, {Mathis}, {Charbonnel},
  \& {Amard}}]{Gallet2017}
{Gallet}, F., {Bolmont}, E., {Mathis}, S., {Charbonnel}, C., \& {Amard}, L.
  2017, \aap, 604, A112

\bibitem[{{Gallet} \& {Bouvier}(2013)}]{Gallet2013}
{Gallet}, F., \& {Bouvier}, J. 2013, \aap, 556, A36

\bibitem[{{Gallet} \& {Bouvier}(2015)}]{Gallet2015}
---. 2015, \aap, 577, A98

\bibitem[{{Goldreich}(1966)}]{Goldreich1966b}
{Goldreich}, P. 1966, \aj, 71, 1

\bibitem[{{Goldreich} \& {Peale}(1966)}]{GoldreichPeale1966}
{Goldreich}, P., \& {Peale}, S. 1966, \aj, 71, 425

\bibitem[{{Goldreich} \& {Soter}(1966)}]{Goldreich1966}
{Goldreich}, P., \& {Soter}, S. 1966, \icarus, 5, 375

\bibitem[{{Greenberg}(2009)}]{Greenberg2009}
{Greenberg}, R. 2009, \apjl, 698, L42

\bibitem[{{Habets} \& {Zwaan}(1989)}]{Habets1989}
{Habets}, G.~M.~H.~J., \& {Zwaan}, C. 1989, \aap, 211, 56

\bibitem[{{Haisch} {et~al.}(2001){Haisch}, {Lada}, \& {Lada}}]{Haisch2001}
{Haisch}, Jr., K.~E., {Lada}, E.~A., \& {Lada}, C.~J. 2001, \apjl, 553, L153

\bibitem[{{Hamers} {et~al.}(2016){Hamers}, {Perets}, \& {Portegies
  Zwart}}]{Hamers2016}
{Hamers}, A.~S., {Perets}, H.~B., \& {Portegies Zwart}, S.~F. 2016, \mnras,
  455, 3180

\bibitem[{{Heller} {et~al.}(2011){Heller}, {Leconte}, \& {Barnes}}]{Heller2011}
{Heller}, R., {Leconte}, J., \& {Barnes}, R. 2011, \aap, 528, A27

\bibitem[{{Herbst} {et~al.}(2001){Herbst}, {Bailer-Jones}, \&
  {Mundt}}]{Herbst2001}
{Herbst}, W., {Bailer-Jones}, C.~A.~L., \& {Mundt}, R. 2001, \apjl, 554, L197

\bibitem[{{Herbst} {et~al.}(2002){Herbst}, {Bailer-Jones}, {Mundt},
  {Meisenheimer}, \& {Wackermann}}]{Herbst2002}
{Herbst}, W., {Bailer-Jones}, C.~A.~L., {Mundt}, R., {Meisenheimer}, K., \&
  {Wackermann}, R. 2002, \aap, 396, 513

\bibitem[{{Howell} {et~al.}(2014){Howell}, {Sobeck}, {Haas}, {Still},
  {Barclay}, {Mullally}, {Troeltzsch}, {Aigrain}, {Bryson}, {Caldwell},
  {Chaplin}, {Cochran}, {Huber}, {Marcy}, {Miglio}, {Najita}, {Smith},
  {Twicken}, \& {Fortney}}]{Howell2014}
{Howell}, S.~B., {Sobeck}, C., {Haas}, M., {et~al.} 2014, \pasp, 126, 398

\bibitem[{Hunter(2007)}]{Hunter2007}
Hunter, J.~D. 2007, Computing In Science \& Engineering, 9, 90

\bibitem[{{Hurley} {et~al.}(2002){Hurley}, {Tout}, \& {Pols}}]{Hurley2002}
{Hurley}, J.~R., {Tout}, C.~A., \& {Pols}, O.~R. 2002, \mnras, 329, 897

\bibitem[{{Hut}(1981)}]{Hut1981}
{Hut}, P. 1981, \aap, 99, 126

\bibitem[{{Irwin} \& {Bouvier}(2009)}]{Irwin2009}
{Irwin}, J., \& {Bouvier}, J. 2009, in IAU Symposium, Vol. 258, The Ages of
  Stars, ed. E.~E. {Mamajek}, D.~R. {Soderblom}, \& R.~F.~G. {Wyse}, 363--374

\bibitem[{{Ivanova} {et~al.}(2005){Ivanova}, {Belczynski}, {Fregeau}, \&
  {Rasio}}]{Ivanova2005}
{Ivanova}, N., {Belczynski}, K., {Fregeau}, J.~M., \& {Rasio}, F.~A. 2005,
  \mnras, 358, 572

\bibitem[{{Jackson} {et~al.}(2009){Jackson}, {Barnes}, \&
  {Greenberg}}]{Jackson2009}
{Jackson}, B., {Barnes}, R., \& {Greenberg}, R. 2009, \apj, 698, 1357

\bibitem[{{Jackson} {et~al.}(2008){Jackson}, {Greenberg}, \&
  {Barnes}}]{Jackson2008}
{Jackson}, B., {Greenberg}, R., \& {Barnes}, R. 2008, \apj, 678, 1396

\bibitem[{{Keppens}(1997)}]{Keppens1997}
{Keppens}, R. 1997, \aap, 318, 275

\bibitem[{{Khaliullin} \& {Khaliullina}(2011)}]{Khaliullin2011}
{Khaliullin}, K.~F., \& {Khaliullina}, A.~I. 2011, \mnras, 411, 2804

\bibitem[{{Leconte} {et~al.}(2010){Leconte}, {Chabrier}, {Baraffe}, \&
  {Levrard}}]{Leconte2010}
{Leconte}, J., {Chabrier}, G., {Baraffe}, I., \& {Levrard}, B. 2010, \aap, 516,
  A64

\bibitem[{{Levato}(1974)}]{Levato1974}
{Levato}, H. 1974, \aap, 35, 259

\bibitem[{{Lurie} {et~al.}(2017){Lurie}, {Vyhmeister}, {Hawley}, {Adilia},
  {Chen}, {Davenport}, {Juri{\'c}}, {Puig-Holzman}, \&
  {Weisenburger}}]{Lurie2017}
{Lurie}, J.~C., {Vyhmeister}, K., {Hawley}, S.~L., {et~al.} 2017, \aj, 154, 250

\bibitem[{{MacGregor} \& {Brenner}(1991)}]{MacGregor1991}
{MacGregor}, K.~B., \& {Brenner}, M. 1991, \apj, 376, 204

\bibitem[{{Mamajek} \& {Hillenbrand}(2008)}]{Mamajek2008}
{Mamajek}, E.~E., \& {Hillenbrand}, L.~A. 2008, \apj, 687, 1264

\bibitem[{{Mardling} \& {Aarseth}(2001)}]{Mardling2001}
{Mardling}, R.~A., \& {Aarseth}, S.~J. 2001, \mnras, 321, 398

\bibitem[{{Martin} {et~al.}(2015){Martin}, {Mazeh}, \&
  {Fabrycky}}]{Martin2015b}
{Martin}, D.~V., {Mazeh}, T., \& {Fabrycky}, D.~C. 2015, \mnras, 453, 3554

\bibitem[{{Mathis}(2015)}]{Mathis2015}
{Mathis}, S. 2015, \aap, 580, L3

\bibitem[{{Matson} {et~al.}(2018){Matson}, {Howell}, \& {Ciardi}}]{Matson2018}
{Matson}, R.~A., {Howell}, S.~B., \& {Ciardi}, D. 2018, arXiv e-prints,
  arXiv:1811.02108

\bibitem[{{Matt} {et~al.}(2015){Matt}, {Brun}, {Baraffe}, {Bouvier}, \&
  {Chabrier}}]{Matt2015}
{Matt}, S.~P., {Brun}, A.~S., {Baraffe}, I., {Bouvier}, J., \& {Chabrier}, G.
  2015, \apjl, 799, L23

\bibitem[{{Matt} {et~al.}(2019){Matt}, {Brun}, {Baraffe}, {Bouvier}, \&
  {Chabrier}}]{Matt2019}
---. 2019, \apjl, 870, L27

\bibitem[{{Mazeh}(2008)}]{Mazeh2008}
{Mazeh}, T. 2008, in EAS Publications Series, Vol.~29, EAS Publications Series,
  ed. M.-J. {Goupil} \& J.-P. {Zahn}, 1--65

\bibitem[{McKinney(2010)}]{Mckinney2010}
McKinney, W. 2010, in Proceedings of the 9th Python in Science Conference, ed.
  S.~van~der Walt \& J.~Millman, 51 -- 56

\bibitem[{{McQuillan} {et~al.}(2014){McQuillan}, {Mazeh}, \&
  {Aigrain}}]{McQuillan2014}
{McQuillan}, A., {Mazeh}, T., \& {Aigrain}, S. 2014, \apjs, 211, 24

\bibitem[{{Meibom} {et~al.}(2015){Meibom}, {Barnes}, {Platais}, {Gilliland},
  {Latham}, \& {Mathieu}}]{Meibom2015}
{Meibom}, S., {Barnes}, S.~A., {Platais}, I., {et~al.} 2015, \nat, 517, 589

\bibitem[{{Meibom} \& {Mathieu}(2005)}]{Meibom2005}
{Meibom}, S., \& {Mathieu}, R.~D. 2005, \apj, 620, 970

\bibitem[{{Meibom} {et~al.}(2006){Meibom}, {Mathieu}, \&
  {Stassun}}]{Meibom2006}
{Meibom}, S., {Mathieu}, R.~D., \& {Stassun}, K.~G. 2006, \apj, 653, 621

\bibitem[{{Mestel}(1968)}]{Mestel1968}
{Mestel}, L. 1968, \mnras, 138, 359

\bibitem[{{Moe} \& {Kratter}(2018)}]{Moe2018}
{Moe}, M., \& {Kratter}, K.~M. 2018, \apj, 854, 44

\bibitem[{{Mu{\~n}oz} \& {Lai}(2015)}]{Munoz2015}
{Mu{\~n}oz}, D.~J., \& {Lai}, D. 2015, Proceedings of the National Academy of
  Science, 112, 9264

\bibitem[{{Murray} \& {Dermott}(1999)}]{Murray1999}
{Murray}, C.~D., \& {Dermott}, S.~F. 1999, {Solar system dynamics}

\bibitem[{{Ogilvie}(2013)}]{Ogilvie2013}
{Ogilvie}, G.~I. 2013, \mnras, 429, 613

\bibitem[{{Ogilvie} \& {Lin}(2007)}]{Ogilvie2007}
{Ogilvie}, G.~I., \& {Lin}, D.~N.~C. 2007, \apj, 661, 1180

\bibitem[{{Orosz} {et~al.}(2012){Orosz}, {Welsh}, {Carter}, {Fabrycky},
  {Cochran}, {Endl}, {Ford}, {Haghighipour}, {MacQueen}, {Mazeh},
  {Sanchis-Ojeda}, {Short}, {Torres}, {Agol}, {Buchhave}, {Doyle}, {Isaacson},
  {Lissauer}, {Marcy}, {Shporer}, {Windmiller}, {Barclay}, {Boss}, {Clarke},
  {Fortney}, {Geary}, {Holman}, {Huber}, {Jenkins}, {Kinemuchi}, {Kruse},
  {Ragozzine}, {Sasselov}, {Still}, {Tenenbaum}, {Uddin}, {Winn}, {Koch}, \&
  {Borucki}}]{Orosz2012}
{Orosz}, J.~A., {Welsh}, W.~F., {Carter}, J.~A., {et~al.} 2012, Science, 337,
  1511

\bibitem[{{Raghavan} {et~al.}(2010){Raghavan}, {McAlister}, {Henry}, {Latham},
  {Marcy}, {Mason}, {Gies}, {White}, \& {ten Brummelaar}}]{Raghavan2010}
{Raghavan}, D., {McAlister}, H.~A., {Henry}, T.~J., {et~al.} 2010, \apjs, 190,
  1

\bibitem[{{Reiners} \& {Mohanty}(2012)}]{Reiners2012}
{Reiners}, A., \& {Mohanty}, S. 2012, \apj, 746, 43

\bibitem[{{Reinhold} {et~al.}(2018){Reinhold}, {Bell}, {Kuszlewicz}, {Hekker},
  \& {Shapiro}}]{Reinhold2018}
{Reinhold}, T., {Bell}, K.~J., {Kuszlewicz}, J., {Hekker}, S., \& {Shapiro},
  A.~I. 2018, arXiv e-prints, arXiv:1810.11250

\bibitem[{{Reinhold} {et~al.}(2013){Reinhold}, {Reiners}, \&
  {Basri}}]{Reinhold2013}
{Reinhold}, T., {Reiners}, A., \& {Basri}, G. 2013, \aap, 560, A4

\bibitem[{{Repetto} \& {Nelemans}(2014)}]{Repetto2014}
{Repetto}, S., \& {Nelemans}, G. 2014, \mnras, 444, 542

\bibitem[{{Ricker} {et~al.}(2014){Ricker}, {Winn}, {Vanderspek}, {Latham},
  {Bakos}, {Bean}, {Berta-Thompson}, {Brown}, {Buchhave}, {Butler}, {Butler},
  {Chaplin}, {Charbonneau}, {Christensen-Dalsgaard}, {Clampin}, {Deming},
  {Doty}, {De Lee}, {Dressing}, {Dunham}, {Endl}, {Fressin}, {Ge}, {Henning},
  {Holman}, {Howard}, {Ida}, {Jenkins}, {Jernigan}, {Johnson}, {Kaltenegger},
  {Kawai}, {Kjeldsen}, {Laughlin}, {Levine}, {Lin}, {Lissauer}, {MacQueen},
  {Marcy}, {McCullough}, {Morton}, {Narita}, {Paegert}, {Palle}, {Pepe},
  {Pepper}, {Quirrenbach}, {Rinehart}, {Sasselov}, {Sato}, {Seager},
  {Sozzetti}, {Stassun}, {Sullivan}, {Szentgyorgyi}, {Torres}, {Udry}, \&
  {Villasenor}}]{Ricker2014}
{Ricker}, G.~R., {Winn}, J.~N., {Vanderspek}, R., {et~al.} 2014, in \procspie,
  Vol. 9143, Space Telescopes and Instrumentation 2014: Optical, Infrared, and
  Millimeter Wave, 914320

\bibitem[{{Rodr{\'{\i}}guez} {et~al.}(2012){Rodr{\'{\i}}guez}, {Callegari},
  {Michtchenko}, \& {Hussmann}}]{Rodriguez2012}
{Rodr{\'{\i}}guez}, A., {Callegari}, N., {Michtchenko}, T.~A., \& {Hussmann},
  H. 2012, \mnras, 427, 2239

\bibitem[{{Rodr{\'{\i}}guez-Ledesma} {et~al.}(2009){Rodr{\'{\i}}guez-Ledesma},
  {Mundt}, \& {Eisl{\"o}ffel}}]{Rodriguez-Ledesma2009}
{Rodr{\'{\i}}guez-Ledesma}, M.~V., {Mundt}, R., \& {Eisl{\"o}ffel}, J. 2009,
  \aap, 502, 883

\bibitem[{{Simonian} {et~al.}(2018){Simonian}, {Pinsonneault}, \&
  {Terndrup}}]{Simonian2018}
{Simonian}, G.~V.~A., {Pinsonneault}, M.~H., \& {Terndrup}, D.~M. 2018, ArXiv
  e-prints, arXiv:1809.02141

\bibitem[{{Skumanich}(1972)}]{Skumanich1972}
{Skumanich}, A. 1972, \apj, 171, 565

\bibitem[{{Stassun} {et~al.}(1999){Stassun}, {Mathieu}, {Mazeh}, \&
  {Vrba}}]{Stassun1999}
{Stassun}, K.~G., {Mathieu}, R.~D., {Mazeh}, T., \& {Vrba}, F.~J. 1999, \aj,
  117, 2941

\bibitem[{{Sullivan} {et~al.}(2015){Sullivan}, {Winn}, {Berta-Thompson},
  {Charbonneau}, {Deming}, {Dressing}, {Latham}, {Levine}, {McCullough},
  {Morton}, {Ricker}, {Vanderspek}, \& {Woods}}]{Sullivan2015}
{Sullivan}, P.~W., {Winn}, J.~N., {Berta-Thompson}, Z.~K., {et~al.} 2015, \apj,
  809, 77

\bibitem[{{Tokovinin} {et~al.}(2006){Tokovinin}, {Thomas}, {Sterzik}, \&
  {Udry}}]{Tokovinin2006}
{Tokovinin}, A., {Thomas}, S., {Sterzik}, M., \& {Udry}, S. 2006, \aap, 450,
  681

\bibitem[{{Torres} {et~al.}(2018){Torres}, {Curtis}, {Vanderburg}, {Kraus}, \&
  {Rizzuto}}]{Torres2018}
{Torres}, G., {Curtis}, J.~L., {Vanderburg}, A., {Kraus}, A.~L., \& {Rizzuto},
  A. 2018, \apj, 866, 67

\bibitem[{van~der Walt {et~al.}(2011)van~der Walt, Colbert, \&
  Varoquaux}]{vanderWalt2011}
van~der Walt, S., Colbert, S.~C., \& Varoquaux, G. 2011, Computing in Science
  Engineering, 13, 22

\bibitem[{{Van Eylen} {et~al.}(2016){Van Eylen}, {Winn}, \&
  {Albrecht}}]{vanEylen2016}
{Van Eylen}, V., {Winn}, J.~N., \& {Albrecht}, S. 2016, \apj, 824, 15

\bibitem[{{van Saders} {et~al.}(2018){van Saders}, {Pinsonneault}, \&
  {Barbieri}}]{vanSaders2018}
{van Saders}, J.~L., {Pinsonneault}, M.~H., \& {Barbieri}, M. 2018, ArXiv
  e-prints, arXiv:1803.04971

\bibitem[{{Verbunt} \& {Zwaan}(1981)}]{Verbunt1981}
{Verbunt}, F., \& {Zwaan}, C. 1981, \aap, 100, L7

\bibitem[{{Witte} \& {Savonije}(2002)}]{Witte2002}
{Witte}, M.~G., \& {Savonije}, G.~J. 2002, \aap, 386, 222

\bibitem[{{Zahn}(1975)}]{Zahn1975}
{Zahn}, J.-P. 1975, \aap, 41, 329

\bibitem[{{Zahn}(1994)}]{Zahn1994}
---. 1994, \aap, 288, 829

\bibitem[{{Zahn}(2008)}]{Zahn2008}
{Zahn}, J.-P. 2008, in EAS Publications Series, Vol.~29, EAS Publications
  Series, ed. M.-J. {Goupil} \& J.-P. {Zahn}, 67--90

\bibitem[{{Zahn} \& {Bouchet}(1989)}]{Zahn1989}
{Zahn}, J.-P., \& {Bouchet}, L. 1989, \aap, 223, 112

\end{thebibliography}

\end{document}